%% file: EWexcess20.latex
\documentclass[a4paper]{article}

\usepackage{amsmath}  % for `\DeclareMathOperator`
\usepackage{bm}
\usepackage{booktabs}
\usepackage{graphicx}
\usepackage{fullpage}
\PassOptionsToPackage{hyphens}{url}\usepackage{hyperref}
\usepackage{longtable}

\newcommand{\mata}{\bm{\alpha}}
\newcommand{\vect}[1]{\ensuremath{\bm{#1}}}
\newcommand{\vo}{\vect{\omega}}

\newcommand{\ttAGE}{\texttt{AGE}}
\newcommand{\ttM}{\texttt{M}}

\begin{document}

\title{\textbf{Excess registered deaths in England and Wales during the COVID-19 pandemic, March 2020 to May 2020}}
\author{Drew M Thomas\\\texttt{<}\href{mailto:dmt107@imperial.ac.uk}{\texttt{dmt107@imperial.ac.uk}}\texttt{>}}
\date{\today}

\maketitle

\begin{abstract}
Official counts of COVID-19 deaths have been criticized for potentially including people who did not die \emph{of} COVID-19 but merely died \emph{with} COVID-19.
I address that critique by fitting a generalized additive model to weekly counts of all deaths registered in England and Wales during the 2010s.
The model produces baseline rates of death registrations expected without the COVID-19 pandemic, and comparing those baselines to recent counts of registered deaths exposes the emergence of excess deaths late in March 2020.
By April's end, England and Wales registered 45,300${} \pm 3200$ excess deaths of adults aged 45+.
Through 22 May, the last day of available all-deaths data, 56,600${} \pm 4400$ were registered (about 53\% of which were of men).
Both the ONS's corresponding count of 43,205 death certificates which mention COVID-19, and the Department of Health and Social Care's count of 33,671 deaths, are appreciably less, implying that their counting methods have underestimated, not overestimated, the pandemic's true death toll.
If underreporting rates have held steady during May, about 59,000 direct and indirect COVID-19 deaths might have been registered through the end of May but not yet publicly reported in full.
\end{abstract}

\tableofcontents

\section{Introduction}

The ongoing COVID-19 pandemic has killed hundreds of thousands of people \cite[p.\ 1]{WHO20}, including tens of thousands in England and hundreds in Wales \cite{DHSC20}.
False negatives and false positives obstruct estimation of exactly how many people COVID-19 has killed; false negatives are people who die of COVID-19 but whose deaths are wrongly attributed to something else, and false positives are people whose deaths are wrongly attributed to COVID-19.
While official COVID-19 death counts boomed, assorted commentators, some well credentialled and some well known, stressed the possibility of false positives inflating the death count.
Andrew Neil, host of BBC Television's \emph{Politics Live} and \emph{The Andrew Neil Show}, former chairman of Sky TV, and chairman of \emph{The Spectator} magazine, tweeted \cite{Neil20}:
\begin{quotation}
Various headlines now hitting multiple websites with variations of: $''$UK death toll due to \#Coronavirus now 766, an increase of 182 in last 24 hours$''$ \\
BUT: what these stats don't tell us us: [sic] \\
How many died FROM the virus? \\
How many died WITH the virus?
\end{quotation}
Forensic pathologist Raquel Fortun, of the University of the Philippines, tweeted \cite{Fortun20}:
\begin{quotation}
So covid-19 kills old people, and those sick of other things. In the Philippines are we really counting right? Or do we just say they're old, or died of something else?[\texttt{THINKING FACE} emoji]
\end{quotation}
Bill Mitchell, Florida-based host of talk show YourVoice\texttrademark \,America, tweeted \cite{Mitchell20a}:
\begin{quotation}
I want to know how many who $''$died from COVID-19$''$ were on ventilators prior to their death? Because if your COVID-19 wasn't bad enough to put you on a ventilator, you died from something else.

Where are the numbers?
\end{quotation}
Mitchell also tweeted \cite{Mitchell20b}:
\begin{quotation}
I say the COVID-19 death counts are BULLSH*T. I say they are loaded with deaths from the flu and other co-morbidities and have been $''$dumped$''$ into the COVID-19 column.

I demand a transparent international standard on what constitutes a COVID-19 death! 

Retweet if you agree.
\end{quotation}
Professional boxer and former professional rugby player Anthony Mundine posted on Facebook \cite{Mundine20}:
\begin{quotation}
Corona virus [sic] is bogus for real ! There [sic] putting fear through media to set an agenda!All [sic] the deaths there [sic] saying that's happening is true but then they add the corona virus [sic] when it was something else to scare everybody \& market the corona ! The world order is happening peeps ! STAY WOKE NOT ASLEEP in what's going on around you ! Mass vaccines will be introduced soon you will see !! To harm \& control you more \& could cause major effects on adults \& children like autism \& even death ! Then they will blame corona!
\end{quotation}
John P.\ A.\ Ioannidis, co-director of METRICS (the Meta-Research Innovation Center at Stanford) and editor-in-chief of the \textit{European Journal of Clinical Investigation}, observed more temperately in an opinion piece for STAT \cite{Ioannidis20} that
\begin{quotation}
[i]n some people who die from viral respiratory pathogens, more than one virus is found upon autopsy and bacteria are often superimposed. A positive test for coronavirus does not mean necessarily that this virus is always primarily responsible for a patient's demise.
\end{quotation}
Pundit Candace Owens tweeted to her two million Twitter followers \cite{Owens20}:
\begin{quotation}
In other words: the death toll we are seeing is ``people that tested positive for the the virus, and are now dead.'' Not, people that have died FROM Coronavirus.

That number, when investigated, will be much lower. Italy determined only 12\% of their death toll was FROM Covid
\end{quotation}
BBC News's main UK Twitter account tweeted to 10.8 million followers \cite{BBCNews20}:
\begin{quotation}
Deaths being reported daily are hospital cases where a person dies with the coronavirus infection in their body

But is the virus causing the death?

It could be the major cause, a contributory factor or simply present when they die of something else 
\end{quotation}
Evidently the question remains open of how much false positives are inflating COVID-19 death counts.

This paper contributes to answering that question with death data from the United Kingdom's Office for National Statistics (ONS).
The basic idea is simple.
Instead of trying to count COVID-19 deaths directly and exclusively, I study how all known deaths, from any cause, have changed over time.
A sufficiently large and rapid increase in COVID-19 deaths should appear as statistically detectable increases in all registered deaths, and increases in those registered deaths would be the same regardless of how many non-COVID-19 deaths were misregistered as COVID-19 deaths.

To try to detect those statistical signals, I fit a statistical model to the ONS's pre-COVID-19 counts of all registered deaths, and generate baseline estimates of how many deaths would be expected absent the pandemic, given historical trends.
I then compare the observed number of registered deaths in recent weeks against those baselines to see whether a statistically significant excess of registered deaths is evident; in the absence of an alternative explanation for them, such excess deaths may be attributed to COVID-19.

\section{Data}

\subsection{Death counts}

In this study I use the ONS's provisional counts of deaths registered in England and Wales each week \cite{ONSweekly20} from week 1 (2--8 January) of 2010 onwards.
Only 11 days out of date when published, those provisional data are the most up-to-date national counts of all deaths.
While the ONS also publishes daily-resolution counts of all deaths registered in England, the daily-resolution counts appear only in the ONS's quarterly death statistics \cite{ONSquarterly20}, which have a far greater months-long publication delay.

%%%%% THIS PARAGRAPH APPLIES ONLY TO RESPIRATORY-DISEASE DEATHS. %%%%%
%The provisional weekly counts extend back to 2010, but I use only the counts from week 2 (4--10 January) of 2014 onwards, because the ONS changed the software it used to code causes of death on 1 January 2014, switching from the Mortality Medical Data System (MMDS) to version 2013 of IRIS \cite{ONSguide19}.
%For week 2 of 2014 the ONS counted respiratory-disease deaths using both pieces of software, with the MMDS producing a count of 1925 and IRIS a count of 1783, a nontrivial discrepancy of 7.4\% (more than 3 times larger than the random week-to-week fluctuations of 2.2\%--2.4\% that one would expect from Poisson statistics).
%I therefore use only the counts from week 2 of 2014 onwards to obtain a more consistent dataset.
%The ONS did change its coding software again on 1 January 2020, from IRIS to MUSE 5.5 \cite{ONSswitch19}.
%However, that switch had a far smaller impact on the respiratory-disease death counts, reducing them by only 0.6\% \cite{ONSswitch19} (far less than the week-to-week random fluctuations of 1.3\% implied by Poisson statistics).

In its provisional weekly counts the ONS decomposes the total counts in several ways.
It tabulates each week's deaths by respiratory disease separately from the all-cause totals (though the counts of respiratory-disease deaths are difficult to integrate with the other death counts, because respiratory-disease deaths overlap only partially with deaths associated with COVID-19 \cite{Caul20}).
The ONS also breaks down the weekly counts by decedents' region of usual residence (North West, Yorkshire and The Humber, Wales, etc.), and by sex and age group.
My analyses use the breakdown by age and sex (appendix \ref{apx:full-age-sex-gam}); because deaths by region are broken down separately, it is not possible to analyze the deaths by age, sex, and region simultaneously.

Before April 2020, the weekly counts used 7 age bands when breaking down deaths by age: age 0, ages 1--14, ages 15--44, ages 45--64, ages 65--74, ages 75--84, and ages 85+.
During the COVID-19 pandemic, however, the ONS updated the counts for 2020 to use a finer breakdown with 20 age bands: age 0, ages 1--4, then 17 contiguous 5-year age bands, and finally ages 90+.
For consistency's sake I aggregate the revised data for 2020 back to the original coarser age banding, reconciling them with the older data.

\subsection{Central England temperatures}

As well as the death counts themselves, I use the Met Office Hadley Centre for Climate Change's time series of daily outdoor minimum temperatures and daily outdoor maximum temperatures in Central England \cite{Parker92,HadCETmin,HadCETmax}.
Ambient, outdoor air temperatures near surface level could plausibly correlate with COVID-19 deaths and deaths in general, and Central England temperatures (CET) are a reasonable proxy for typical outdoor air temperatures in England and Wales generally, because outdoor air temperatures correlate reasonably well even at distances of hundreds of kilometres \cite{Parker92,North11} and different regions of England and Wales show similar year-to-year temperature changes \cite[pp.\ 25--58]{Jenkins09}.

I generate 6 weekly aggregates of CET (table \ref{tab:CET-weeklies}) from the daily records as potential inputs to my models of death counts.
Two (\texttt{TMIN} and \texttt{TMAX}) of the aggregates are a particular week's extreme temperatures; a third, \texttt{TMID}, is the average temperature in a week; two more (\texttt{TSD} and \texttt{TRAN}) measure intra-week variability in temperature, and the last, \texttt{TMDI} measures the week-on-week change in temperature.

I also reformat the temperature observations, which the Hadley Centre presents as integers with units of tenths of a degree Celsius, by dividing them each by ten.
The temperature-based parts of my analyses therefore use temperatures in units of degrees Celsius.

\begin{table}
\centering
\begin{tabular}{ccl}
\toprule
aggregate & & definition \\
\midrule
\texttt{TMIN} & & lowest daily minimum temperature \\
\texttt{TMAX} & & highest daily maximum temperature \\
\texttt{TSD} & & standard deviation of a week's minimum and maximum temperatures \\
\texttt{TMID} & & mean of a week's minimum and maximum temperatures \\
\texttt{TMDI} & & a week's \texttt{TMID} minus the previous week's \texttt{TMID} \\
\texttt{TRAN} & & \texttt{TMAX} minus \texttt{TMIN} \\
\bottomrule
\end{tabular}
\caption{\label{tab:CET-weeklies}6 weekly aggregates of daily Central England temperature measurements.}
\end{table}

\subsection{Air-quality indices}

In the United Kingdom, Defra (the Department for Environment, Food and Rural Affairs) publishes a Daily Air Quality Index (DAQI) at individual air-monitoring sites \cite{Defrasites20} and for regions of the UK \cite{Defraregions20}.
The DAQI is the highest of 5 indices of different pollutants: ozone, nitrogen dioxide, sulphur dioxide, PM$_{2.5}$, and PM$_{10}$.
Research has correlated short-term changes in the concentrations of all 5 of these pollutants with short-term changes in mortality in developed countries \cite{Shah13,Shah15,Atkinson14,Chiusolo11,Williams14} and in China \cite{Lai13,Shang13}, and so I generate weekly aggregates of the DAQI to use as potential correlates of registered deaths.

Defra publishes a DAQI for each of 16 regions: one represents all of Northern Ireland, 4 represent Scotland (``Central Scotland'', ``Highland'', ``North East Scotland'', ``Scottish Borders''), two Wales (``North Wales'' and ``South Wales''), and 9 England.
I compute a national DAQI as the simple, unweighted mean of the 11 English and Welsh regional DAQIs.
In theory, it may be better to use a weighted mean, such as a mean weighted by regions' populations, but Defra's DAQI regions do not comport to any single official scheme, which is an obstacle to precisely determining appropriate weights, and with the regional DAQIs all positively correlated with each other (during the 2010--2019 decade, the Pearson product-moment correlations between regions' DAQIs ranged from +0.167 to +0.737, with a mean of +0.407), a weighted mean is unlikely to be a great improvement.

I compute 3 weekly AQI aggregates, \texttt{AQIMIN}, \texttt{AQIMAX}, and \texttt{AQIMID}, as the lowest, highest, and mean of a week's 7 national DAQIs.
Unlike temperature, I define no aggregates representing AQI's variability within or between weeks.
Variation in temperature might cause thermal stress and hence mortality in itself; if people have adapted to the current temperature, then a change in temperature, even towards a more moderate temperature, could induce stress.
There is no analogy to fluctuations in air quality.
A lower (D)AQI should always tend to accompany lower mortality, and there is no basis to expect changes in (D)AQI to trigger or prevent deaths in their own right.

\section{Modelling}

\subsection{Death-registrations data summary and formal model specification}

\begin{figure}
\centering
\includegraphics[width=0.99\textwidth]{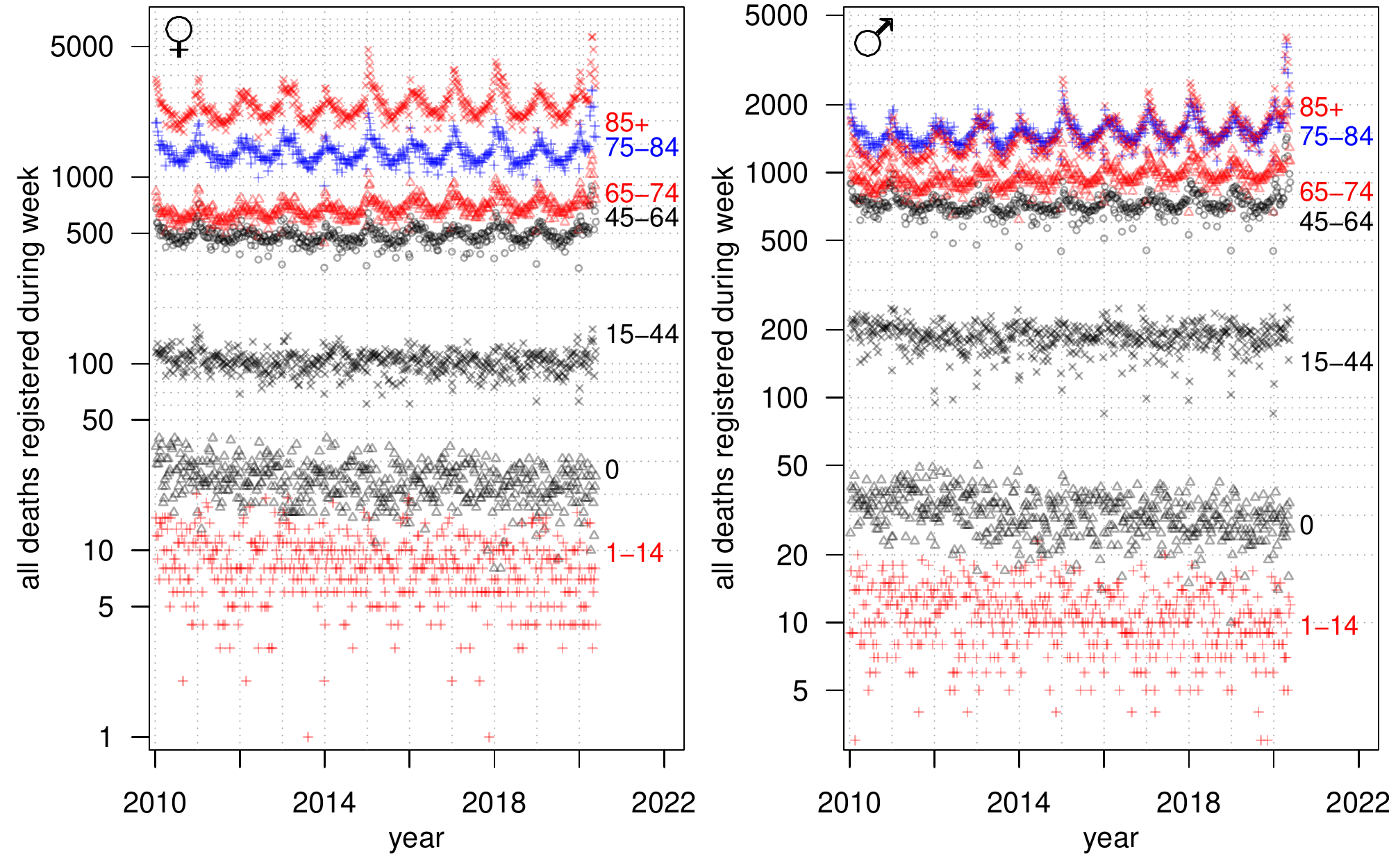}
\caption{Weekly registered deaths in England and Wales, 2010--2020, by age band and sex (female deaths in left panel and male deaths in right panel). In-plot labels state age bands.}
\label{fig:all-deaths-by-all-age-bands}
\end{figure}

Figure \ref{fig:all-deaths-by-all-age-bands} shows the number of deaths registered in England and Wales by week.
Deaths of middle-aged and older adults evince an annual cycle which becomes more pronounced with increasing age.
The cycle peaks around January every year and sinks to a minimum every summer.
Secular (non-cyclical) trends are also visible.
Registered deaths of children less than 15 years old gradually decreased during the 2010s, while those of adults aged 65--74 gradually increased.
A model is necessary to adjust for both seasonality and secular trends when calculating baselines for death rates during the COVID-19 pandemic.

I fit a generalized additive model (GAM) to all of the death counts from week 1 of 2010 through week 9 of 2020, the week ending 28 February 2020.
When fitting the GAM I exclude March 2020 and following months to avoid inflating the GAM's baselines by contaminating them with COVID-19-associated deaths.

The model assumes that each registered-deaths count has a Poisson distribution, and uses the natural-logarithm function as its link function. 
My chosen GAM implementation is the \texttt{gamm} function in R's \texttt{mgcv} package, estimating smoothing parameters with the \texttt{REML} method.

Using \texttt{gamm} I fit the GAM
\begin{align}
\textnormal{deaths } & \textnormal{registered for age band } \ttAGE \textnormal{ and sex } \ttM \textnormal{ in the week ending } \texttt{WKEDAY} \sim \nonumber \\
\textnormal{Po}
& ( \exp( \, \mata_{\ttM, \ttAGE} + \vect{\beta}_{\texttt{SH}}
  + \vect{\gamma}_{\texttt{LSH}}
  + \vo_1{\texttt{ROY}} + \vo_2{\texttt{ESTR}} + \vo_3{\texttt{XMAS}}
  + \vo_4{\texttt{LROY}} + \vo_5{\texttt{LESTR}} + \vo_6{\texttt{LXMAS}}
    \nonumber \\
& \quad\quad\, + \mathbf{s}_{1, \ttM, \ttAGE}(\texttt{WK})
  + \mathbf{s}_{2, \ttM, \ttAGE}(\texttt{WKEDAY})
    \nonumber \\
& \quad\quad + \mathbf{s}_{3, \ttM, \ttAGE}(\texttt{TMID})
  + \mathbf{s}_{4, \ttAGE}(\texttt{TMDI})
  + \bm{\zeta}_{1, \ttM, \ttAGE} \texttt{TRAN}
  + \bm{\zeta}_{2, \ttAGE} \texttt{AQIMIN} \, ))
%  + \mathbf{s}_{5,\ttAGE}(\texttt{TRAN})
%  + \mathbf{s}_{6,\ttAGE}(\texttt{AQIMIN}) \, ))
\label{eq:all-deaths-gam}
\end{align}
where
$\mata$ accounts for the differences in average deaths across ages and sexes;
$\vect{\beta}$, $\vect{\gamma}$, and $\vo$ account for holidays, when register offices may be closed;
and the multipliers represented by $\bm{\zeta}$ and smooths represented by $\mathbf{s}$ account for how death registrations change with the annual cycle, any secular trend over the decade, CET, and AQI.

Specifically: $\mata_{\ttM, \ttAGE}$ is the natural logarithm of mean weekly deaths for sex \ttM\ and age band \ttAGE;
\ttAGE\ is an integer index from 1 through 7, with age 0 having index 1, ages 1--14 having index 2, etc.;
\ttM\ is 0 for females and 1 for males;
$\vect{\beta}$ represents the change in registrations if the week contains a particular secular holiday;
\texttt{SH} represents which (if any) secular holiday the week contains (an integer index from 1 through 5, representing no holiday, the first Monday in May,\footnote{With the exception of the year 2020, for which the government moved the first-Monday-in-May holiday to 8 May for the $75^{\textnormal{\footnotesize th}}$ anniversary of VE Day.} the last Monday in August, the spring bank holiday, and New Year's Day);
$\vect{\gamma}$ represents the change in registrations if the \emph{previous} week contained a particular secular holiday;
\texttt{LSH} is \texttt{SH} with a 1-week lag (i.e.\ which secular holiday, if any, the previous week contained);
$\vo$ represents the change in registrations for each Easter or Christmas holiday the week contains (or the previous week contained);
\texttt{ROY}, \texttt{ESTR} and \texttt{XMAS} are the number of Royal Family-related holidays, Easter holidays (Good Friday or Easter Monday), and the number of Christmas holidays (Christmas Day and/or Boxing Day) respectively in the week;
\texttt{LROY}, \texttt{LESTR}, and \texttt{LXMAS} are \texttt{ROY}, \texttt{ESTR}, and \texttt{XMAS} with a 1-week lag (i.e.\ how many Royal/Easter/Christmas holidays were in the previous week);
\texttt{WK} is the week of the year (an integer between 1 and 53);
\texttt{WKEDAY} is the date when the week ended, represented as the number of days since 1 January 2010;
\texttt{TMID}, \texttt{TMDI}, and \texttt{TRAN} are 3 of the temperature variables mentioned above;
\texttt{AQIMIN} is the weekly minimum AQI;
$\bm{\zeta}$ represents how death registrations vary with \texttt{TRAN} and \texttt{AQIMIN}; and
$\mathbf{s}$ represents age-dependent (and sometimes sex-dependent) smooths of week-of-year, of date, and temperature variables.

To decide whether to use smooths ($\mathbf{s}$) or a simple linear multiplier ($\bm{\zeta}$) to relate a variable to death registrations, and which age-dependent relationships had to be split into multiple relationships for different sexes, I fitted a maximally flexible GAM (i.e.\ one where all non-categorical variables were fitted with sex-and-age-specific smooths).
I reviewed the fitted relationships from that GAM to discern where I could simplify the GAM (by collapsing multiple sex-specific smooths of a variable into one sex-independent smooth, or by replacing a potentially nonlinear smooth with a simple linear relationship) without notable underfitting.
Appendix \ref{apx:full-age-sex-gam} discusses that review, and equation \ref{eq:all-deaths-gam} is the review's outcome: the simplified model I use as my main GAM in this paper.

\subsection{Fitted model of death registrations}

\begin{table}
\centering
\small
\begin{tabular}{lrcr}
\toprule
& \multicolumn{3}{c}{parameter values} \\
\cmidrule{2-4}
factor & on natural-log.\ scale & & on original scale \\
\midrule
girls aged 0 ($\mata_{0,1}$)        & $3.167 \pm 0.068$ & & $  24 \pm  2$ \\
girls aged 1--14 ($\mata_{0,2}$)    & $1.998 \pm 0.101$ & & $   7 \pm  1$ \\
females aged 15--44 ($\mata_{0,3}$) & $4.566 \pm 0.031$ & & $  96 \pm  3$ \\
women aged 45--64 ($\mata_{0,4}$)   & $6.145 \pm 0.018$ & & $ 467 \pm  8$ \\
women aged 65--74 ($\mata_{0,5}$)   & $6.500 \pm 0.015$ & & $ 665 \pm 10$ \\
women aged 75--84 ($\mata_{0,6}$)   & $7.191 \pm 0.011$ & & $1328 \pm 15$ \\
women aged 85+ ($\mata_{0,7}$)      & $7.761 \pm 0.009$ & & $2347 \pm 22$ \\
boys aged 0 ($\mata_{1,1}$)       & $3.284 \pm 0.059$ & & $  27 \pm  2$ \\
boys aged 1--14 ($\mata_{1,2}$)   & $2.212 \pm 0.095$ & & $   9 \pm  1$ \\
males aged 15--44 ($\mata_{1,3}$) & $5.171 \pm 0.030$ & & $ 176 \pm  5$ \\
men aged 45--64 ($\mata_{1,4}$)   & $6.554 \pm 0.016$ & & $ 702 \pm 12$ \\
men aged 65--74 ($\mata_{1,5}$)   & $6.847 \pm 0.014$ & & $ 941 \pm 13$ \\
men aged 75--84 ($\mata_{1,6}$)   & $7.301 \pm 0.011$ & & $1481 \pm 17$ \\
men aged 85+ ($\mata_{1,7}$)      & $7.244 \pm 0.010$ & & $1400 \pm 14$ \\
\midrule
first Monday in May during week ($\vect{\beta}_2 - \vect{\beta}_1$)
	& $-0.144 \pm 0.004$ & & $(86.6 \pm 0.3)$\% \\
last Monday in August during week ($\vect{\beta}_3 - \vect{\beta}_1$)
	& $-0.120 \pm 0.004$ & & $(88.7 \pm 0.4)$\% \\
New Year's Day during week ($\vect{\beta}_4 - \vect{\beta}_1$)
	& $-0.153 \pm 0.012$ & & $(85.8 \pm 1.0)$\% \\
spring bank holiday during week ($\vect{\beta}_5 - \vect{\beta}_1$)
	& $-0.133 \pm 0.004$ & & $(87.5 \pm 0.4)$\% \\
Royal Family-related holidays during week ($\vo_1$)
	& $-0.108 \pm 0.009$ & & $(89.7 \pm 0.8)$\% \\
Easter holidays during week ($\vo_2$)
	& $-0.143 \pm 0.003$ & & $(86.6 \pm 0.3)$\% \\
Christmas holidays during week ($\vo_3$)
	& $-0.207 \pm 0.002$ & & $(81.3 \pm 0.2)$\% \\
\midrule
first Monday in May during prev.\ week ($\vect{\gamma}_2 - \vect{\gamma}_1$)
	& $0.024 \pm 0.004$ & & $(102.4 \pm 0.4)$\% \\
last Monday in August during prev.\ week ($\vect{\gamma}_3 - \vect{\gamma}_1$)
	& $0.028 \pm 0.004$ & & $(102.8 \pm 0.4)$\% \\
New Year's Day during prev.\ week ($\vect{\gamma}_4 - \vect{\gamma}_1$)
	& $0.079 \pm 0.004$ & & $(108.2 \pm 0.4)$\% \\
spring bank holiday during prev.\ week ($\vect{\gamma}_5 - \vect{\gamma}_1$)
	& $0.032 \pm 0.004$ & & $(103.2 \pm 0.4)$\% \\
Royal Family-related holidays during prev.\ week ($\vo_4$)
	& $0.094 \pm 0.008$ & & $(109.8 \pm 0.9)$\% \\
Easter holidays during prev.\ week ($\vo_5$)
	& $0.097 \pm 0.003$ & & $(110.1 \pm 0.3)$\% \\
Christmas holidays during prev.\ week ($\vo_6$)
	& $0.056 \pm 0.006$ & & $(105.7 \pm 0.6)$\% \\
\bottomrule
\end{tabular}
\caption{\label{tab:all-deaths-gam-1} Average weekly death registrations by sex and age band, and average holiday-associated changes in weekly death registrations, as estimated by the GAM in eq.\ \ref{eq:all-deaths-gam}. ``$\pm$'' symbols denote approximate standard errors.}
\end{table}

\begin{figure}
\centering
\includegraphics[width=0.99\textwidth]{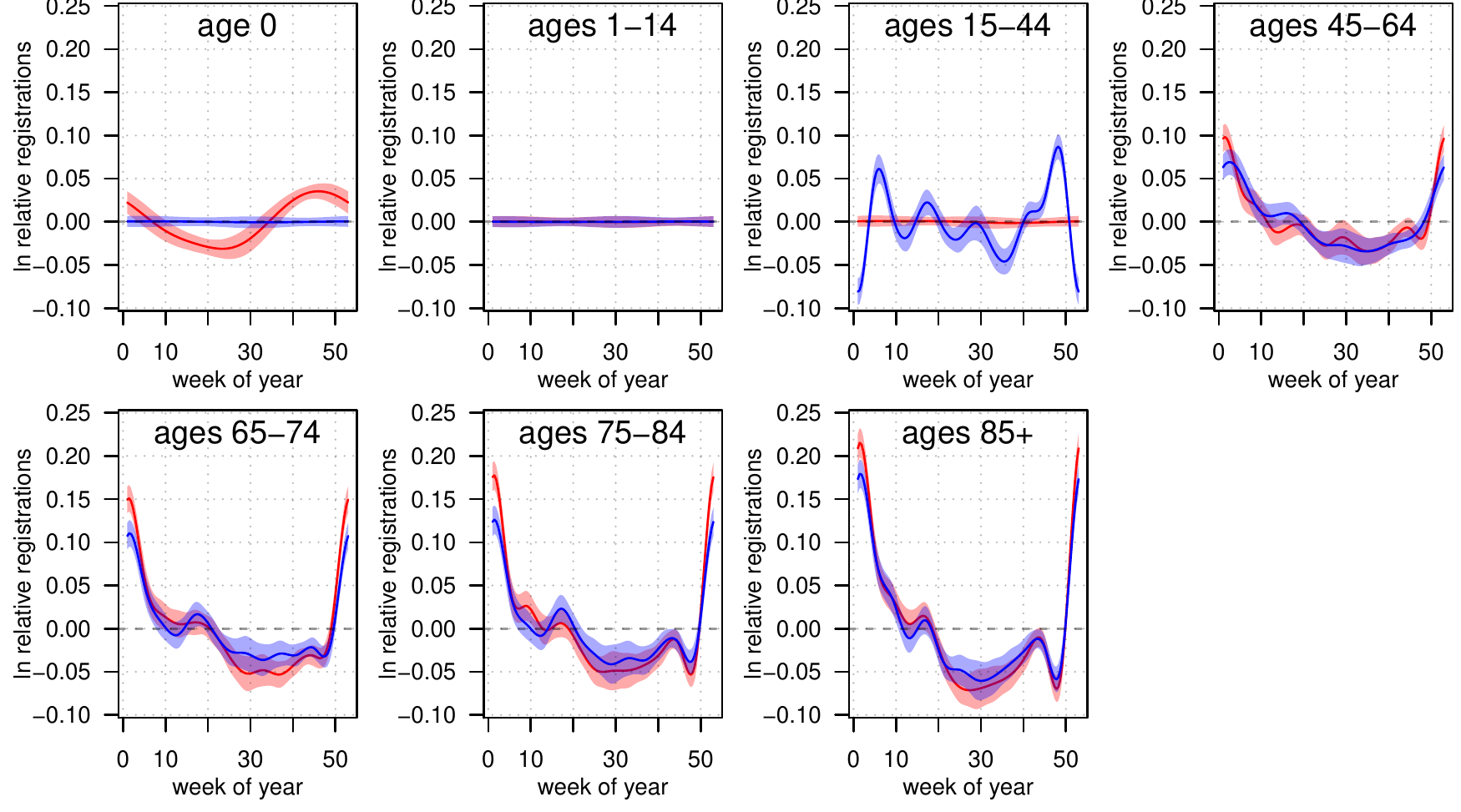}
\caption{Annual cycles in registered deaths estimated as relative registered deaths by week of year and age band, measured on the GAM's natural-log.\ scale. Confidence bands represent $\pm 2$ standard errors around each estimated cycle. Age 0 is omitted because the GAM found no statistically significant cycle.}
\label{fig:annual-cycle-by-age-band}
\end{figure}

\begin{figure}
\centering
\includegraphics[width=0.99\textwidth]{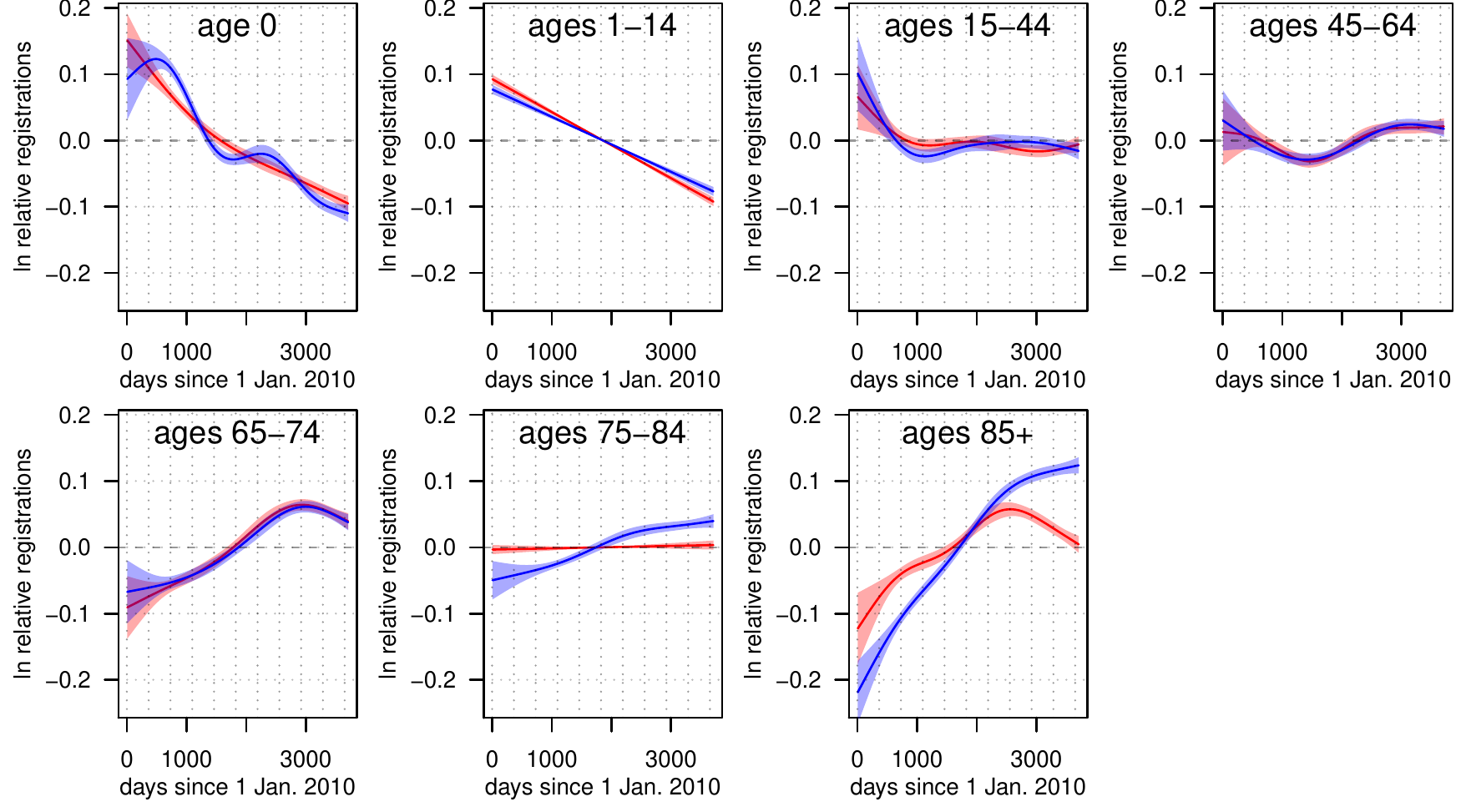}
\caption{Secular trends in registered deaths by age band, measured on the GAM's natural-log.\ scale. Confidence bands extend $\pm 2$ standard errors around fitted trends. The $x$ axis is the date represented as the number of days since 1 Jan.\ 2010. Vertical dotted gridlines denote New Year's Day, 2010--2020.}
\label{fig:secular-trends-by-age-band}
\end{figure}

\begin{figure}
\centering
\includegraphics[width=0.32\textwidth]{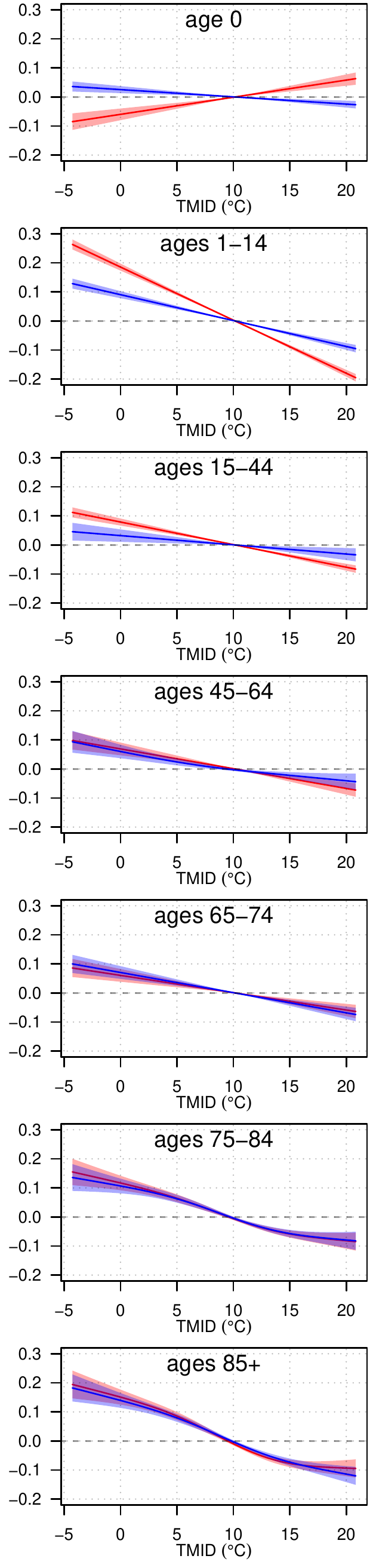}
\includegraphics[width=0.32\textwidth]{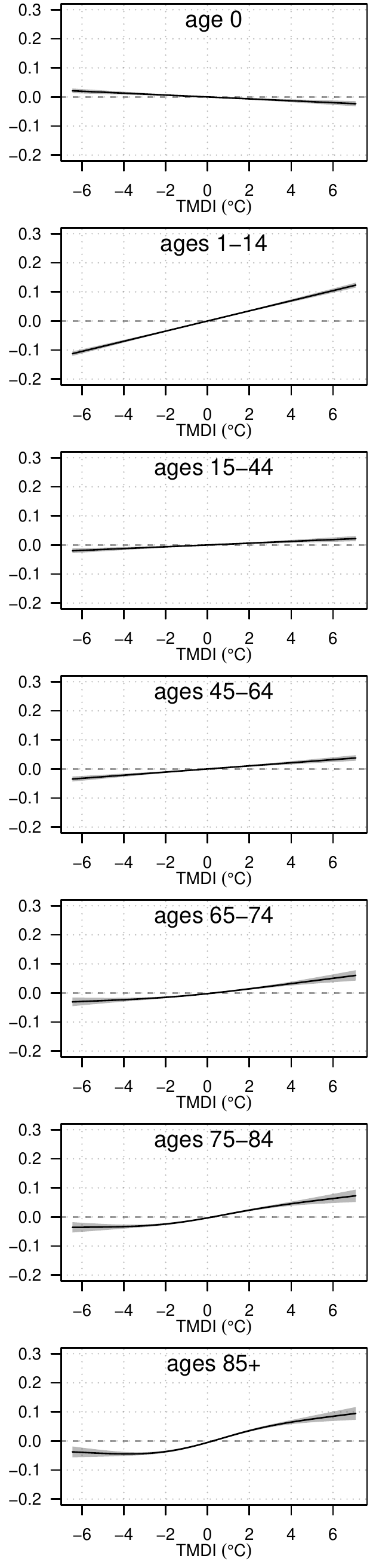}
\includegraphics[width=0.32\textwidth]{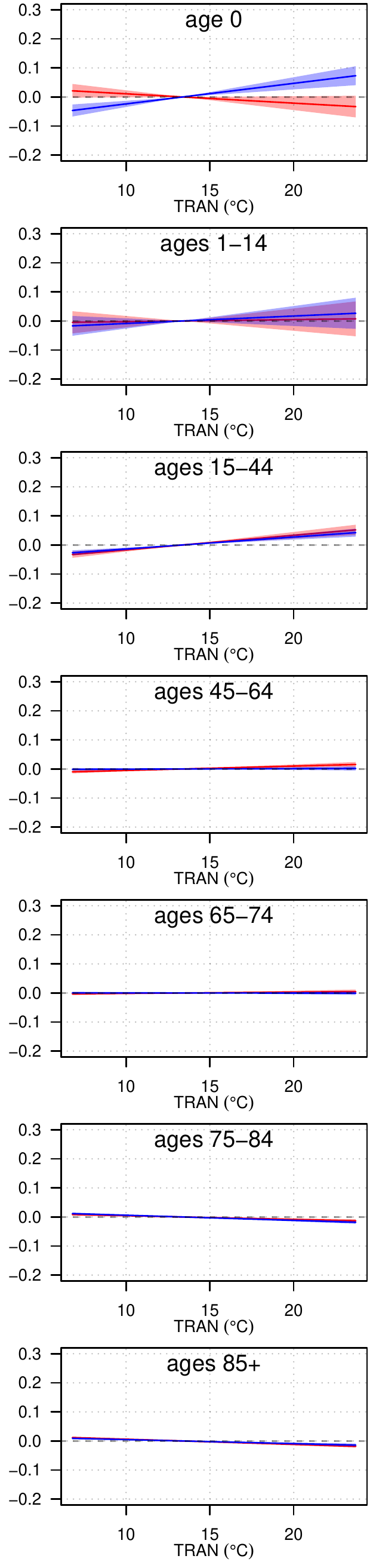}
\caption{Registered deaths as a function of \texttt{TMID} (left), \texttt{TMDI} (centre), and \texttt{TRAN} (right) by age band, measured on the GAM's natural-log.\ scale. Confidence bands extend $\pm 2$ standard errors around each fit. Female-specific, male-specific, and sex-independent fits are red, blue, and grey respectively.}
\label{fig:temperature-by-age-band}
\end{figure}

\begin{figure}
\centering
\includegraphics[width=0.99\textwidth]{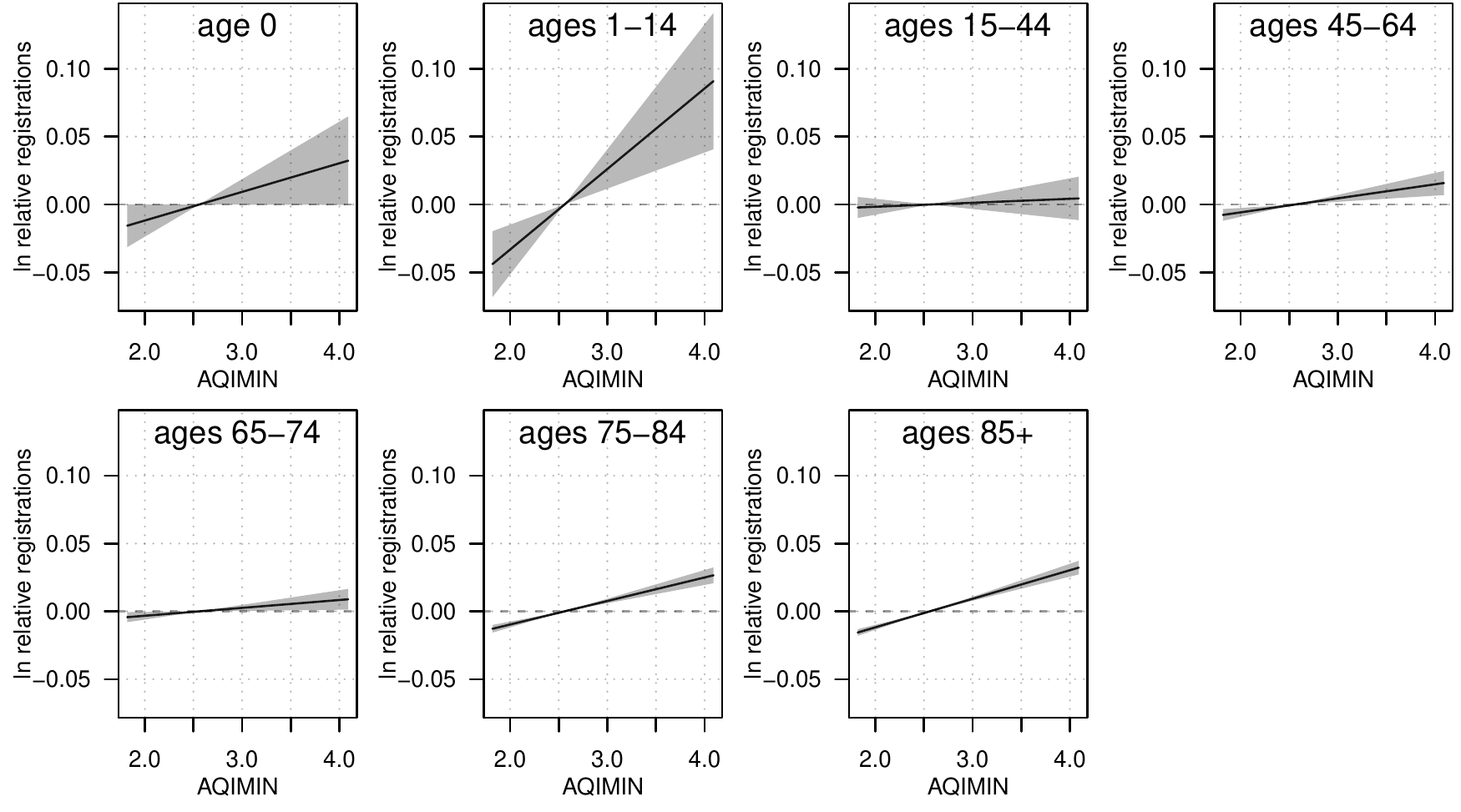}
\caption{Registered deaths as a function of \texttt{AQIMIN} by age band, measured on the GAM's natural-log.\ scale. Confidence bands extend $\pm 2$ standard errors around each fit.}
\label{fig:aqi-by-age-band}
\end{figure}

Fitting the model to the time series of weekly death registrations produces a range of findings about the relationships registered deaths have with sex, age, week of year, CET, and air quality (table \ref{tab:all-deaths-gam-1}, figs. \ref{fig:annual-cycle-by-age-band}, \ref{fig:secular-trends-by-age-band}, \ref{fig:temperature-by-age-band}, and \ref{fig:aqi-by-age-band}) under normal circumstances.

Beginning with the results for age and sex, registered deaths have a J-shaped relationship to age regardless of sex (table \ref{tab:all-deaths-gam-1}).
Children aged 1--14 suffer the fewest registered deaths, with more deaths registered in a typical week among younger children (infants less than 12 months old) and adolescents and adults.
There are more registered deaths of males than of females in every age band, except in the age-85+ age band, where the sex difference reverses and women average 68\% more death registrations.

The results further reveal that, on average, when a public holiday occurs during a week, death registrations are 10\%--19\% lower in that week, but 2\%--11\% higher the week after (table \ref{tab:all-deaths-gam-1}).
Those estimates comport with a naive back-of-envelope estimate that a one-day holiday might reduce a week's registrations by one fifth to one seventh, because register offices would effectively lose one of the week's 5--7 working days, but could partially compensate by registering slightly more deaths the week after.\footnote{That the compensation is only partial implies at least one of the following: (\emph{i}) fewer people die in holiday weeks, (\emph{ii}) register offices take multiple weeks to clear a typical backlog of yet-to-be-registered holiday-week deaths, and/or (\emph{iii}) register offices fail to register some holiday-week deaths entirely.}
Not all holidays are equal.
Over the decade of the 2010s, only 10\% fewer deaths were registered on Royal Family-related holidays, but 19\% fewer were registered for each Christmas holiday during a week --- and because most Christmas weeks include two public holidays (Christmas Day and Boxing Day), the model implies that in a typical Christmas week, 34\% fewer deaths would be registered.

Figure \ref{fig:annual-cycle-by-age-band} plots $\mathbf{s}_{1, \ttM, \ttAGE}(\texttt{WK})$, the annual cycles the model estimates for each age band.
The GAM finds no statistically significant annual cycles in pre-adolescent children's registered deaths, except for a minor cycle for infant girls.
In teenage boys and men aged up to 45, a somewhat inscrutable cycle emerges which oscillates almost from month to month.
Then, in all of the older age bands, death registrations have a broadly U-shaped relationship to time of year, and this U-shaped cycle is marginally stronger in women.
Death registrations peak in the winter, trend downwards during spring, and reach a nadir in July and August, before increasing as winter approaches.
That overall cycle also includes some subtler features: registrations tend to drop in November, between October's rise and December's rapid winter ramp-up, and the V-shaped mini-cycle around weeks 10--15 tends to be mirrored by an opposing mini-cycle around weeks 15--20.

The holiday multipliers and annual-cycle smooths account for the calendar cycle in registered deaths.
Figure \ref{fig:secular-trends-by-age-band} moves on to $\mathbf{s}_{2, \ttM, \ttAGE}(\texttt{WKEDAY})$, which estimates the secular trends in registered deaths over the decade.
That addresses the long-term trends in deaths due to slowly changing variables, most obviously England and Wales's growing population.
Registered deaths of young children consistently decreased throughout the 2010s, but trends were less encouraging at ages 15+.
From 2010 through 2012, death registrations of those aged 15--64 steadily decreased, but changed little thereafter.
Finally, there is some hetereogeneity among the trends for people aged 65+ (the trend for adults, especially women, aged 75--84 is curiously muted relative to the trends for 65--74-year-olds and those at least 85 years old), but the graphs strongly suggest that they died in increasing numbers from 2010 until 2018.
Those increases cannot be put down to winter influenza because the annual-cycle smooths already account for causes of death with a seasonal cycle (which include the average winter epidemic), and these long-term secular-trend smooths average out the impact of any one worse-than-average winter flu epidemic.

3 more sets of smooths capture the relationship between CET and registered deaths (fig.\ \ref{fig:temperature-by-age-band}).
The 3 temperature variables the GAM uses are \texttt{TMID}, which captures how hot each week was on average; \texttt{TMDI}, which captures how much hotter a week was than the week before it; and \texttt{TRAN}, which indexes how variable temperature was during a week.
The temperature during a week and inter-week shifts in temperature prove more important than intra-week temperature variability; \texttt{TMID} and \texttt{TMDI} show stronger relationships with registered deaths than \texttt{TRAN}.

\texttt{TMID} generally has a negative relationship with death registrations; mortality appears to be lower in hot weeks, and that relationship is stronger in older adults than younger adults.
\texttt{TMDI}, however, tends to have a positive relationship with death registrations, suggesting that mortality is higher when a week is warmer than the week before.
\texttt{TRAN} is unusual in showing significantly different relationships at different ages: a high/wide \texttt{TRAN} is associated with more registered deaths of those aged 15--44 and \emph{fewer} registered deaths of those aged 75+.
The changing sign of the relationship between \texttt{TRAN} and death registrations across different ages is arguably anomalous, but since the relationship is consistently modest (and virtually nonexistent in those aged 1--14 or 65--74) the sign inconsistency would be a mild anomaly in practical terms.

The remaining pieces of the GAM are the age-dependent fits of \texttt{AQIMIN} (fig.\ \ref{fig:aqi-by-age-band}).
\texttt{AQIMIN} correlates positively with registered deaths, sometimes statistically significantly, which is consistent with earlier epidemiological studies.
When a week's minimum national DAQI is elevated (i.e.\ when average national air-pollution levels are elevated every day for a week), slightly more deaths of (pre)school-age children, middle-aged adults, and retirement-age adults are registered.

\begin{figure}
\centering
\includegraphics[width=0.99\textwidth]{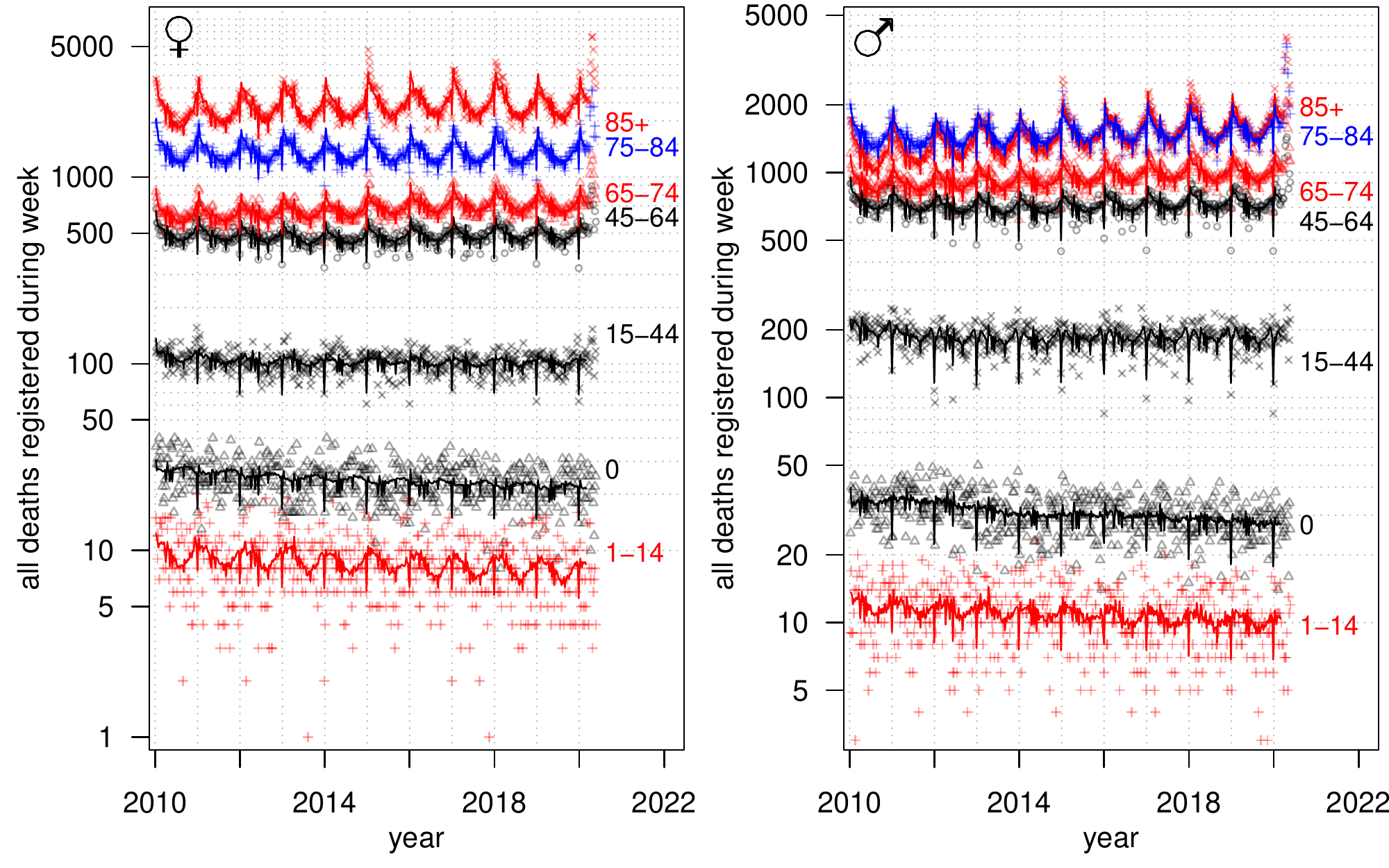}
\caption{Weekly registered deaths, by age band and sex, with GAM fits superimposed as solid curves. In-plot labels state age bands; left and right panels plot female and male deaths respectively.}
\label{fig:all-deaths-by-all-age-bands-with-gam}
\end{figure}

After studying the GAM's individual components, the natural next step is to review the retrospective baselines that the whole GAM produces.
Figure \ref{fig:all-deaths-by-all-age-bands-with-gam} plots the GAM's retrospective baselines against all of the original data.
The fits for each age band successfully reproduce each band's overall rate of registered deaths, and each band's overall secular trend during the 2010s.
The assumption that death registrations are Poisson distributed implies that on the logarithmic-scale plot, the death counts should appear less scattered around higher baselines, and that prediction is borne out; visually, there is a lot of vertical scatter for children (where the baselines are lower), and less vertical scatter for older adults (where the baselines are higher).

\begin{figure}
\centering
\includegraphics[width=0.99\textwidth]{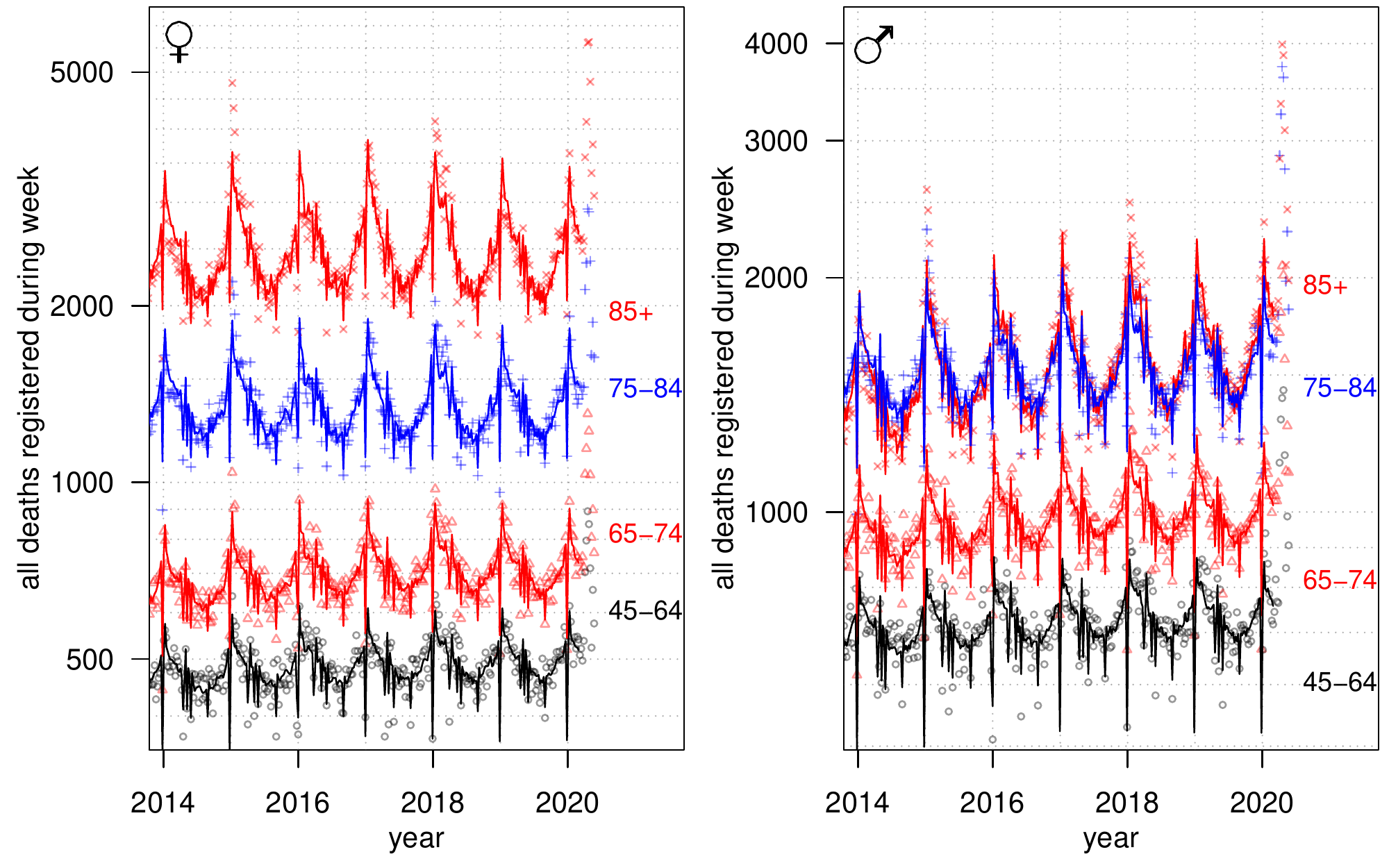}
\caption{Weekly registered deaths in recent years, for each older age band, with GAM fits superimposed as solid curves. In-plot labels state age bands; left and right panels plot female and male deaths respectively.}
\label{fig:all-deaths-by-old-age-bands-with-gam}
\end{figure}

\begin{figure}
\centering
\includegraphics[width=0.99\textwidth]{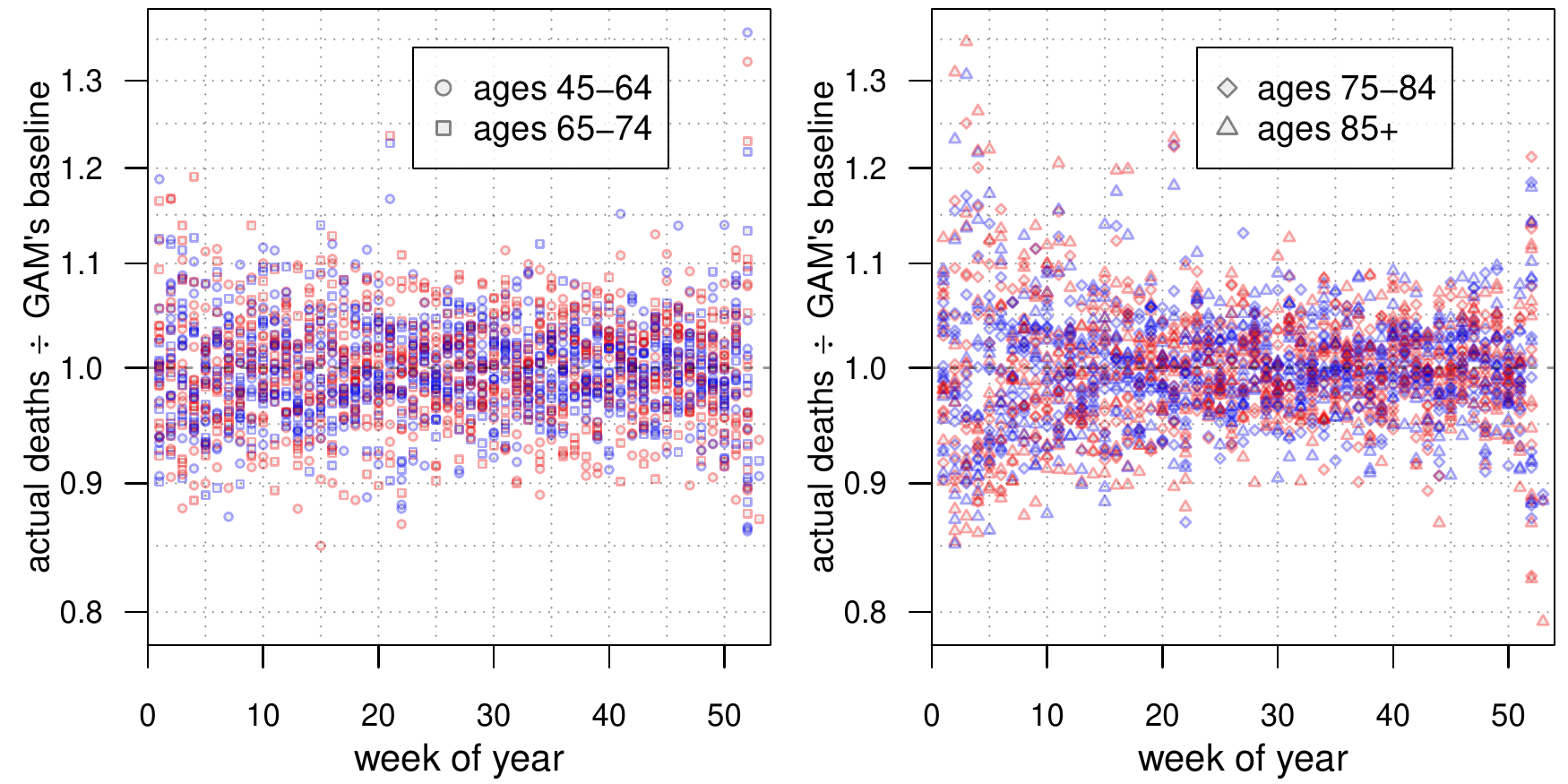}
\caption{Actual weekly registered deaths divided by the GAM's expected weekly registered deaths, for every combination of (older) age band, sex (female observations in red and male observations in blue), and week in the years 2010 through 2019. Ratios under 1 represent GAM overestimates and ratios over 1 underestimates. Red and circles represent female deaths; blue and diamonds represent male deaths.}
\label{fig:older-relative-gam-errors}
\end{figure}

Focusing on the most-recent data and fits for the older age bands (fig.\ \ref{fig:all-deaths-by-old-age-bands-with-gam}) demonstrates that the model fails to precisely match every winter peak in registered old-age deaths.
Checking this more systematically by plotting the GAM's relative errors for older adults by week of year (fig.\ \ref{fig:older-relative-gam-errors}) confirms that the GAM has particular trouble fitting weeks 52--53 and 1--6.
The most obvious explanation is that epidemics of respiratory viruses like influenza and RSV (respiratory syncytial virus) each winter, and concomitant peaks in secondary diseases such as pneumonia, are a major influence on old-age excess mortality in the winter, and the sizes of those epidemics are somewhat random and difficult to predict \cite{HallNaqvi19,Fleming08}.
While respiratory-infection deaths do correlate with the weather and the season, it remains the case that average temperatures, air quality, and the time of year do not fully account for all variation in deaths from respiratory infections, so the GAM cannot account fully for idiosyncratically small or large winter epidemics.
One must bear this limitation in mind if using the GAM to produce baselines for winter weeks, but it is unlikely to interfere much with estimating excess COVID-19 deaths because virtually all UK COVID-19 deaths have occurred in week 11 or later of 2020, after the worst of the RSV/influenza season.

\subsection{Rationales for modelling choices}

One could fit many slight variations on eq.\ \ref{eq:all-deaths-gam}, so I justify some of my modelling choices with my understanding of the data-generating process.
A primary concern informing my modelling is balancing parsimony against the introduction of bias by oversimplifying and underfitting my model.

Parsimony leads me to minimize how many interactions among variables I include in the GAM.
Crudely, the GAM takes 7 quantities as inputs: age, sex, public-holiday status, week of the year (\texttt{WK}), overall date, temperature, and air pollution.
Thus there are $2^7 - 7 - 1 = 120$ broad kinds of interaction the GAM could include in principle, but ultimately I use only a few: the interaction of age, and, where appropriate, sex, with each of \texttt{WK}, date, temperature, and air pollution.

I chose those interactions with age and sex because the results make clear that they are necessary.
Figures \ref{fig:annual-cycle-by-age-band}, \ref{fig:secular-trends-by-age-band}, \ref{fig:temperature-by-age-band}, and \ref{fig:aqi-by-age-band}, along with appendix \ref{apx:full-age-sex-gam}, make plain that the relationships death registrations have with \texttt{WK}, date, temperature, and air pollution differ by age and sometimes sex.
By contrast, there is strong theoretical reason not to have age or sex interact with holiday status.
Death registrations dip on public holidays (and then rebound) because many register offices close on public holidays.
Closed register offices register no deaths, regardless of the age and sex of the deceased.
As such, public holidays should have the same effect on death registrations regardless of age and sex, and trying to account for different effects according to age and sex is pointless (if not actively harmful, inasmuch as it wastes information in the data on superfluous parameters).
I checked that theoretical expectation with a test GAM by adding (linear, not smoothed) age-holiday interaction terms to a GAM and refitting it to the death-registrations data.
Most of the 84 newly added interaction terms were statistically insignificant ($p > 0.1$), and 6 of the 14 original holiday-status terms became statistically insignificant, with 5 of the 14 implausibly becoming positive (i.e.\ implying that death registrations increased on average on those holiday weeks).
Those results support the dubiousness of putting interactions between demographic variables and holiday status into a model of death registrations.

I exclude all 3-way and higher-order interactions from the model because such many-way interactions are difficult (albeit not impossible) to justify a priori, are less readily interpretable than two-way interactions and single regressors' relationships, and threaten to exhaust the information in the dataset.
As an example of the last risk, for each age band and sex, there are only 10 observations of deaths registered on the last Monday in August, because that holiday occurs only 10 times in the period the ONS data cover.
There would therefore be only limited information to fit any 3-way interaction between age or sex, holiday status, and any other variable.

The remaining interactions the GAM could include are two-way interactions not involving age or sex.

I exclude two-way interactions between non-age, non-sex variables and AQI because there is minimal basis to expect them to be practically or statistically significant.
AQI's age-dependent relationships to death registrations tend to be weak; in every adult age band, the highest observed values of \texttt{AQIMIN} accompany death-registration rates at most 3\% higher than average, and that remains so in a more flexible GAM where AQI's age-dependent relationships may differ by sex (appendix \ref{apx:full-age-sex-gam}).
Prior research gives no reason to expect AQI's interactions with other variables to reveal a stronger relationship to registered deaths, except perhaps AQI-temperature interactions, but my experiments with fitting GAMs including \texttt{AQIMIN}-temperature interactions had unimpressive results.

I exclude the non-age, non-AQI two-way interactions because the non-age, non-AQI quantities are all closely related: \texttt{WK} and holiday status are direct functions of the date, and average temperatures have a roughly sinusoidal relationship to \texttt{WK} and date.

Parsimony also informs the maximum number of degrees of freedom (DOF) I allow \texttt{gamm} to use for each of the \ttAGE-dependent smooths.
I limit every smooth of \texttt{TMID}, \texttt{TMDI}, \texttt{TRAN}, and \texttt{AQIMIN} to 3 DOF, since it is implausible that those variables' age-specific relationships to registered deaths are much more complex than S-shaped relationships, which 3 DOF suffice to represent.
Each age band's \texttt{WKEDAY} smooth can use at most 5 DOF, which permits one inflection point per two years in fitted secular death-registrations trend --- enough flexibility to capture multiple shifts in the trend over time, but not so much flexibility as to erroneously absorb individual years' seasonal cycles.
Finally, unlike the other smooths, I force each age band's \texttt{WK} smooth to be cyclic, since \texttt{WK} smooths represent annual cycles.
I limit them to 12 DOF each, granting each \texttt{WK} smooth approximately monthly resolution.
Allotting many more DOF to the \texttt{WK} smooths would risk having them absorb some of the effect of public holidays which fall in the same week each year.

Moving on from matters of parsimony, the residuals of a preliminary fit of the main GAM have slight autocorrelation for the young age bands and medium-sized autocorrelation for old-age adults.
So I fit my model assuming AR(1) residuals within each combination of year and age band.\footnote{Ideally I would fit the model with AR(1) residuals within just each age band, but that proves computationally difficult for \texttt{gamm}. Fortunately, the autocorrelation is not overwhelming (the final model estimates $\phi = 0.3$), so slicing the data into year-long chunks when fitting the AR(1) residual structure does not unduly compromise the model.}

\subsection{Auxiliary models for imputing CET}

The Hadley Centre publishes its daily CET data only for complete months, so temperature data for the current month (June at the time of writing) are unavailable.
That is irrelevant when fitting the GAM, because I fit the GAM only to death counts for 2010 through 28 February 2020, over which period complete CET data are already available.
However, the lack of June data shall become an obstacle when the ONS begins reporting death counts for weeks 23--27 and the GAM has to generate baselines for those weeks, because the necessary current-month temperature data are missing.

Anticipating that problem, I fit additional simple GAMs to the known temperature data, GAMs which I can later use to impute \texttt{TMID}, \texttt{TMDI}, and \texttt{TRAN} for the current month.
%Imputing those variables' values for this month enables me to estimate baselines for this month with the main GAM.

I fit a separate imputation GAM for all of the temperature variables except \texttt{TRAN} (because \texttt{TRAN} is the difference of \texttt{TMAX} and \texttt{TMIN}, so imputing \texttt{TMIN} and \texttt{TMAX} with GAMs suffices to impute \texttt{TRAN} without fitting a specific \texttt{TRAN} GAM).
Every GAM includes a linear date trend, to exploit any consistent secular trends over the 2010--2020 period, and a cyclic smooth of \texttt{WK}, to account for annual temperature cycles.
They also include smooths of the temperature variables \texttt{TMIN}, \texttt{TMAX}, \texttt{TSD}, \texttt{TMID}, and \texttt{TMDI} with a 1-week lag, allowing the smooths only 3 DOF, enough to fit an S-shaped relationship.
Finally, for parsimony, I manually prune extraneous regressors from each GAM and refit them: ultimately, I impute \texttt{TMID}, \texttt{TMDI}, and \texttt{TRAN} as functions of lagged \texttt{TMID} and lagged \texttt{TSD}.
The imputation GAMs are therefore analogous to regularized vector autoregression models, but gently nonlinearized and with a seasonal cycle also fitted.

%This imputation process introduces some error, but the incorporation of seasonality and the relatively short period of imputation (at most 4 weeks if the current month is long, and only 2 weeks in the current study) limits it.
%A sensitivity analysis in the next section finds that the GAM's baselines might be wrong by at most about 10\% due to the error arising from imputing temperatures.
A sensitivity analysis for weeks 18--19, presented in the previous version of this preprint when May temperature data were unavailable, found that the GAM's original May baselines might be wrong by at most about 12\% for adults due to the error arising from imputing temperatures.
Fortunately, as of early June, all necessary temperature data have been published, and I do no imputation of temperature data, so my current results have no imputation error.

\section{Estimation of excess registered deaths}

\begin{figure}
\centering
\includegraphics[width=0.99\textwidth]{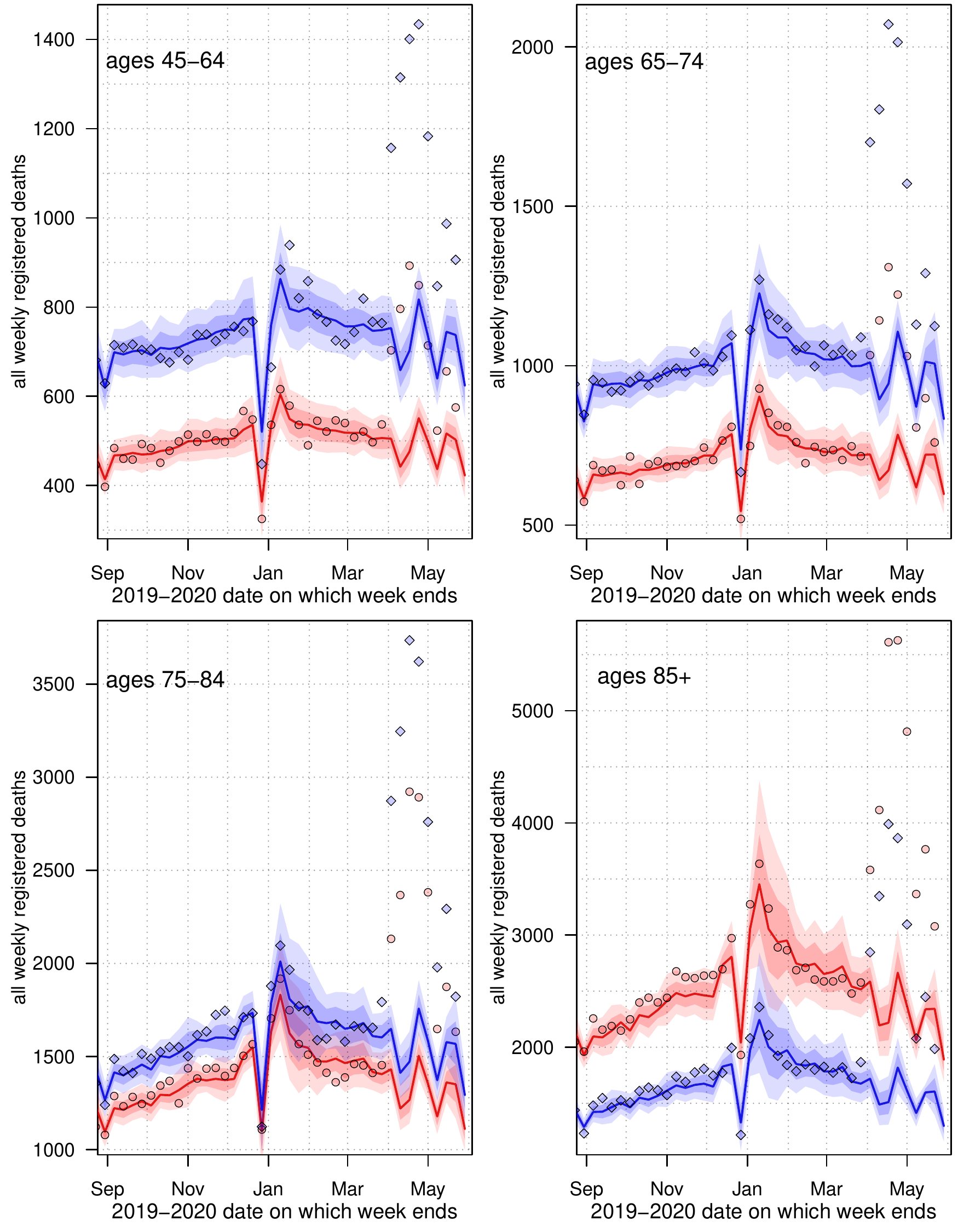}
\caption{Registered deaths of women (``$\circ$'') and of men (``$\diamond$'') compared to the GAM's baselines (solid lines) in older age bands, after August 2019's public holiday. Inner, darker bands cover $\pm 1$ standard error (SE), estimated as the GAM's root-mean-squared log.-scale error in 2010--2019 for the combination of age band and week of year. Outer, lighter bands are 95\% confidence intervals covering $\pm 1.96$ SEs.}
\label{fig:older-close-ups}
\end{figure}

I use the fitted GAM to generate baselines for each age band during this year, and compare the ONS's recent counts to the baselines.
Close-up visual comparisons of the ONS's death counts to the GAM's baselines for older ages are in figure \ref{fig:older-close-ups}.
The observations, plotted as individual circles for women and as individual diamonds for men, run through week 21 of 2020.

The ONS reports 108 registered deaths ``where COVID-19 was mentioned on the death certificate'' in weeks 11 and 12 of this year, but registered deaths in those weeks were nonetheless close to the baselines in all of the older age bands.
In week 13, death registrations increased on week 12 and were mostly above their baselines; moreover, for men aged 75 and up, registered deaths in week 13 were above the baselines' 95\% confidence intervals.
In week 14, death registrations accelerated greatly, producing a statistically significant excess of deaths in each of the older age bands.
Simple visual examination of the plots reveals about 600, 1000, 2000, and 2000 excess registered deaths in age bands 45--64, 65--74, 75--84 and 85+ respectively in week 14, for a total of approximately 5600 excess deaths.
The numbers of excess deaths in each of weeks 15--17 were greater still.

\begin{table}
\centering
\footnotesize
\begin{tabular}{crrrrr}
\toprule
& & \multicolumn{3}{c}{all registered deaths in age band} \\
\cmidrule{3-5}
ages & wk. & GAM baseline & obs'd & excess & ONS C19 \\
\midrule
\input{excess_registrations.latex}
\bottomrule
\end{tabular}
\caption{\label{tab:excess-registrations}Baseline versus observed counts of all deaths, and ONS counts of COVID-19-associated deaths, registered in England and Wales in weeks 12--21 of 2020. ``$\pm$'' symbols denote approximate standard errors. Under-15s' results omitted due to minimal C19 deaths (ONS-documented or excess).}
\end{table}

Table \ref{tab:excess-registrations} collects, for each of weeks 12 through 21 and every adolescent and adult age band, the precise baselines from the GAM, the actual number of registered deaths, the implied number of excess registered deaths, and the ONS's report of COVID-19-associated deaths.
The precise numbers reveal that despite a modest number of their death certificates mentioning COVID-19, the number of excess registered deaths was negligible for adolescents and young adults even into April, although excess registered deaths among 15--44-year-olds achieved statistical significance in week 15, and maintained significance in weeks 16 and 17.
More dramatically, among people aged 45 and up, there were
$83 \pm 436$,
$625 \pm 398$,
$5651 \pm 417$,
$9181 \pm 525$, and
12,656${}\pm 581$
excess registered deaths in weeks 12 through 16 respectively.\footnote{To obtain standard errors of the all-ages weekly totals, I do not sum the individual standard errors in quadrature, but rather sum them in the direct and conventional fashion, lest a given week's errors across age bands be highly correlated.}
After excess registered deaths in that age range peaked in week 16, they declined through weeks 17, 18, and 19 to 10,520${}\pm 495$, $7693 \pm 361$, and $3785 \pm 362$ respectively.
Week 20 brought a slight uptick to $4343 \pm 310$ deaths, followed in week 21 by a dip to $2047 \pm 544$ deaths.

\begin{figure}
\centering
\includegraphics[width=0.99\textwidth]{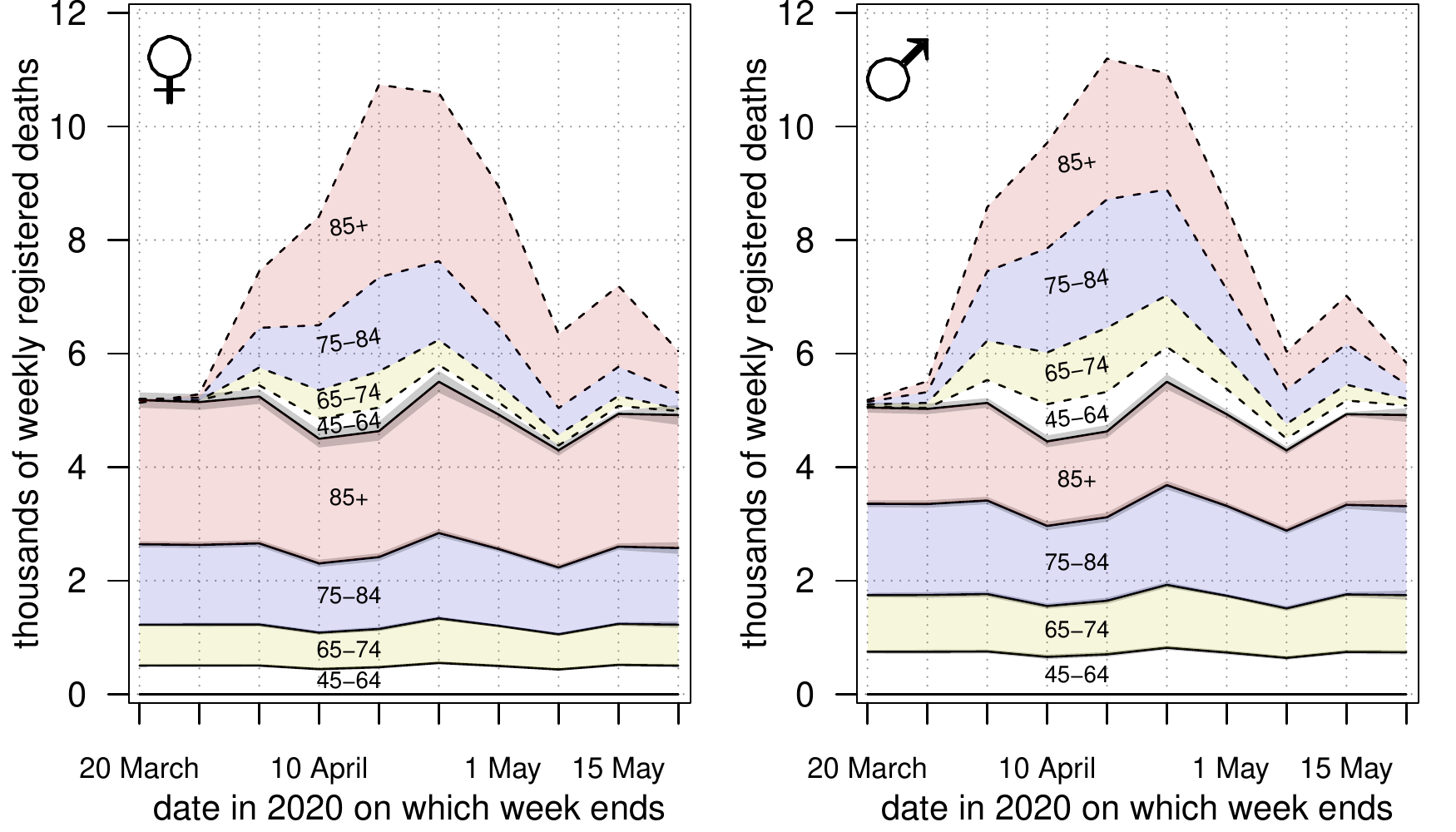}
\caption{Stacked area charts of recent registered deaths (women's in the left panel, men's in the right panel) in older age bands, baselines (areas between adjacent solid lines, with approximate standard errors shown as translucent grey bands around solid lines) and excesses (areas between a dashed line and an adjacent line).}
\label{fig:stacked-older-deaths}
\end{figure}

Summing over weeks 12--21 by calendar month, and pro-rating deaths in weeks that cross month boundaries, produces approximate monthly estimates of $3937 \pm 1072$, 41,386${}\pm 2089$, and 11,274${}\pm 1268$ excess registered deaths of adults aged 45+ in March, April, and the first 22 days of May respectively.
Setting aside May as incomplete, 45,323${} \pm 3161$ excess deaths were registered through the end of April.
Considering all of weeks 12--21, 56,597${} \pm 4429$ excess deaths were registered through 22 May.
Sex-specific data and baselines (appendix \ref{apx:age-week-sex-results}) enable an estimate of the sex distribution of those excess deaths: about 29,744, or 53\%, were of men.
Decomposing the results by age as well as sex belies that near-even split: men made up about 61\% of excess deaths at ages 45--84, but only about 42\% of excess deaths at ages 85+.

Figure \ref{fig:stacked-older-deaths} illustrates the emergence of excess deaths in recent weeks differently, graphically stacking the older age bands' weekly death counts by sex to make them comparable on one scale.
It demonstrates how much total (registered-)death rates increased among older men and older women.
Crudely, in weeks 15 and 16 total registered deaths doubled in each older age band, so COVID-19 roughly doubled the hazard rate of death for people aged 45+.
Closer inspection shows that the increase in risk itself increased with age: in weeks 16 and 17 excess registered deaths of the 85+ visibly outnumbered the baseline death count, but the ratio of excess registered deaths to baseline was somewhat less at ages 45--64.
The diagram also shows that excess registered deaths were not only concentrated in the oldest of the old, but that their excess mortality accelerated the most.
In later weeks, registered deaths of women remained concentrated in the 85+ age band; this is consistent with the old-age population sex ratio being skewed towards women, and disproportionate circulation of SARS-CoV-2 in care homes inhabited mostly by the old \cite{MacAskillGrey20,GreyMacAskill20,GreyMcNeill20,Booth20}.

Comparing the excess-deaths estimates to the ONS's counts of COVID-19-associated deaths suggests that the ONS numbers undercounted people who died, directly or indirectly, because of COVID-19, and that the undercounting was worse in April than in May.
The ONS reported 102, 531, 3432, 6139, 8655, 8134, 5983, 3889, 3776, 2559 ``[d]eaths where COVID-19 was mentioned on the death certificate'' among people aged 45+ in weeks 12--21 respectively.
The first two counts (102 in week 12 and 531 in week 13) do not differ significantly from the relatively noisy GAM-based excess-death estimates, but the subsequent ONS counts are only 61\%, 67\%, 68\%, 77\%, 78\%, 103\%, 87\%, and 125\% respectively of the GAM-based excess-death estimates.
Contrary to the commentators quoted in this paper's introduction, counting the known deaths of those who died with COVID-19/SARS-CoV-2 has tended to \emph{under}estimate how many have died from COVID-19, not \emph{over}estimate.

\section{Conclusion}

This study addresses the concern that COVID-19 has been blamed for too many deaths, an error which could occur if official COVID-19 death counts include too many false positives.
To test for that putative error, I develop a statistical model of historical rates of registered deaths, and compare the baselines that model produces to the actual number of deaths registered during the COVID-19 pandemic.

The comparison gives only a weak statistical signal of excess deaths in week 12 of 2020, but reveals that in weeks 13 through 21 --- i.e.\ from 21 March through 22 May --- of 2020, England and Wales registered 56,769${}\pm 4931$ excess deaths of adolescents and adults, versus the 43,582 registered deaths the ONS reported ``where COVID-19 was mentioned on the death certificate''.
COVID-19 appears to have been blamed for too few deaths, not too many, at least over those 9 weeks.
The true number of registered deaths for which COVID-19 is (directly or indirectly) responsible is likely to be (30 $\pm$ 11)\% higher than the number of registered deaths where the death certificate mentions the disease.

The UK's Department of Health and Social Care provides more up-to-date counts of COVID-19-associated deaths than the ONS, though (until 29 April) their counts of such deaths in England included only people who died in NHS-commissioned medical facilities.
The Department counted 2053 people with COVID-19 who died in England and 81 who died in Wales during week 21, for a total of 2134 deaths \cite{DHSC20}.
During that week, therefore, about $((44 + 241 + 155 + 536 + 1115) \div 2053) = 1.02$ excess deaths were registered for each Department-reported, COVID-19-associated death.
If the same ratio obtains for the 2388 deaths the Department has reported for 23 May through 31 May \cite{DHSC20}, the true number of COVID-19 deaths in England and Wales from 23 May through 31 May may be closer to 2436 (2388 multiplied by 1.02).
Adding that to the estimate of 56,597 excess registered deaths in the 9 weeks leading up to 23 May, the ultimate number of COVID-19 deaths registered (even if not all recognized as COVID-19 deaths) in England and Wales through May's end could be about 59,033, if deaths have been underreported at a steady rate since mid-May.

My work has multiple limitations.
Although it investigates one means of avoiding false-positive errors, it is susceptible to false-negative errors.
My estimates of excess registered deaths can capture only \emph{net} deaths due to the COVID-19 pandemic.
If the response to the pandemic saves some Britons' lives (perhaps by reducing air pollution \cite{Burke20,Chen20} or motor accidents or attacks of non-COVID-19 infectious diseases) while the pandemic itself kills thousands, my approach would detect only the net effect of both taken together, which would underestimate the \emph{gross} number of deaths due to COVID-19.
On the other hand, if the response to the pandemic causes more Britons to die (theoretically possible if the lockdown induces fatal stress or prevents people from obtaining medical attention), my approach would conflate people killed by the response with people killed directly by the pandemic.

The geographical scope of my analysis is quite tightly bounded, confined exclusively to England and Wales, not the entire United Kingdom (although data exist in principle to extend it to the entire UK), nor a wider range of countries.

A last complication is mortality displacement, also known as ``harvesting'' \cite{McMichael98}, which is the phenomenon of a fatal condition killing people only slightly earlier than they otherwise would have died.
In the specific case of COVID-19, it could be that even if all people who die with COVID-19 die because of COVID-19, some of them would not have lived much longer even if they had never caught COVID-19.
If so, and if one blames COVID-19 only for killing those people whose lives it shortened by years, and not for killing people whose lives it shortened by days or weeks, then one can interpret my excess-deaths estimates as overestimates of people killed by COVID-19.
Recent research \cite{Hanlon20} and actuarial commentary \cite{Edwards20} suggest that such mortality displacement is minor, but longer-term data are necessary to exclude it entirely.
If England and Wales suppress the spread of SARS-CoV-2 early in the summer, the weeks immediately after (or at least between waves of) the pandemic might have fewer deaths than the baselines, which would suggest mortality displacement: COVID-19 would have killed in the spring people who would otherwise have died in the summer.\footnote{Unusually low death rates immediately after an acute phase of the pandemic might also reflect an ongoing stay-at-home order, if many people remain at home even when COVID-19 deaths are low, and staying at home turns out to prevent a substantial proportion of non-COVID-19 deaths.}
Also possible, perhaps even probable, is the opposite outcome --- excess deaths could persist beyond the pandemic's acute phase, as they did, for example, after London's Great Smog of 1952 \cite{Bell01}.

\section{Corrigenda}

The 23 April version of this preprint (\textit{i}) erroneously gave the reference to the DH\&SC's daily death counts \cite{DHSC20} an access date of 21 April rather than the correct 22 April; (\textit{ii}) referred in its conclusion to ``the subsequent 8466 hospital deaths from 11 April through 20 April'' when the end of the relevant period was actually 22 April, not 20 April; (\textit{iii}) gave the ratio of excess registered deaths to reported in-hospital deaths in week 15 as 1.91 rather than 1.90; and (\textit{iv}) used an outdated caption for table 3 where ``weeks 12--14'' should have been ``weeks 12--15''.

The 28 April version of this preprint included an outdated table 3 (of expected versus observed counts) where the baselines and standard errors and hence excess-death estimates came from the older GAM without AQI as an input.
Its conclusion also asserted that this study ``estimates excess registered deaths only as a function of time and approximate age'', omitting the additional (albeit also time-dependent) inputs of ambient temperature and AQI.

The 3 May version of this preprint wrongly opened section 4 with the clause ``After imputing temperatures for April,'', a phrase mistakenly retained from the previous version.

\appendix

\section{Guiding modelling choices with a trial full age-sex GAM}

\label{apx:full-age-sex-gam}

The ONS publishes provisional weekly counts of deaths broken down in two ways, namely by region alone, and by sex and age band.
There is no 3-way breakdown by sex and age band \emph{and} region, which forces a choice of which breakdown to use when fitting a model: the region-only breakdown versus the age-and-sex breakdown.
Because age is the variable with the most relevance to COVID-19 mortality, and because the age-by-sex breakdown is somewhat finer-grained than the regional breakdown, I choose to base my modelling on the age-by-sex breakdown rather than the regional breakdown.

The next modelling choice is whether, for each numerical input (\texttt{WK}, \texttt{WKEDAY}, \texttt{TMID}, \texttt{TMDI}, \texttt{TRAN}, \texttt{AQIMIN}) in the model, the main GAM should (\textit{i}) use only one smooth of the input, (\textit{ii}) use one smooth for all males and a second smooth for all females, (\textit{iii}) use a different smooth for each age band but not for different sexes, or (\textit{iv}) use a different smooth for every combination of age and sex.
The last option would be the obvious choice if the available data and computer-processing time were infinite, but finite data and CPU time imply that a less complex model (i.e.\ one of options \textit{i}, \textit{ii}, and \textit{iii}) may be most appropriate.

To decide the best kind of fit for each numerical input, I fit the most complex kind of model --- one with a different smooth for every combination of age, sex, and numerical input --- as a trial, and compare its smooths across ages and sexes to ascertain where fitting different smooths for different ages/sexes is necessary, or merely introducing unneeded complexity.
I also check the nonlinearity of each input's fitted smooths in the full model, to discern whether nonlinear smooths are necessary.
Where it seems reasonable to simplify by fitting a linear relationship rather than a more-general nonlinear smooth, by using the same smooth for males and females, and/or by using the same smooth across all age bands, I make the simplification in the main baseline-generating GAM I use in the main analyses and text.

\begin{figure}
\centering
\includegraphics[width=0.99\textwidth]{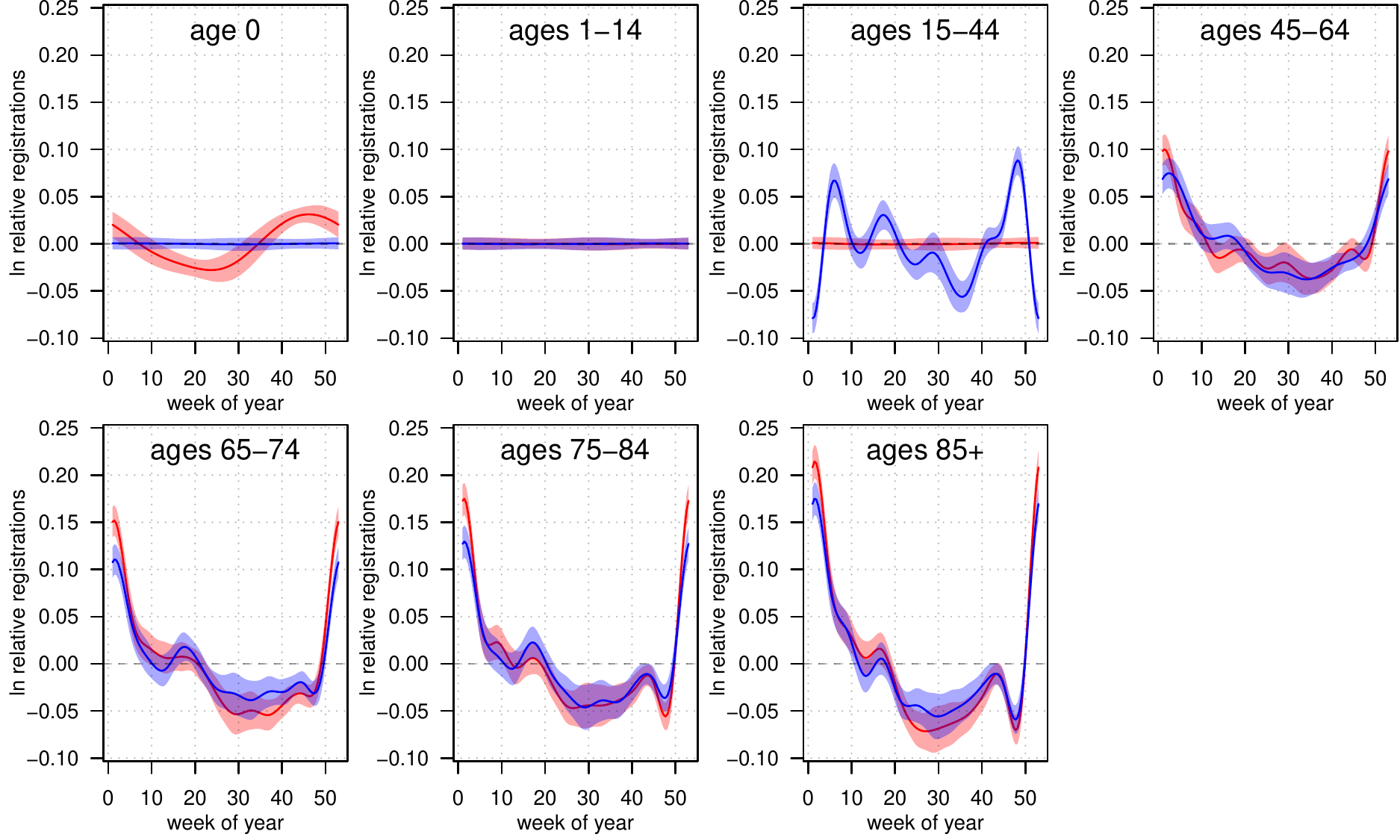}
\caption{Annual cycles in registered deaths estimated as relative registered deaths by week of year and age band and sex (red denoting females and blue males), measured on the GAM's natural-log.\ scale. Confidence bands represent $\pm 2$ standard errors around each estimated cycle.}
\label{fig:age-sex-annual-cycle}
\end{figure}

Figure \ref{fig:age-sex-annual-cycle} presents the fitted smooths that relate \texttt{WK}, the week of the year, to registered deaths for every age-sex combination.
Death registrations have no relationship to \texttt{WK} for children aged 1--14, but at other ages annual cycles are apparent.
Annual cycles are plainest among middle-aged and older adults, becoming stronger in older age bands.
Registered deaths of retirement-age women and of retirement-age men have qualitatively similar annual cycles, but the cycle is systematically slightly stronger in women.
That feature of the data, along with the female-specific relationship between registered deaths and \texttt{WK} in infants, and the male-specific relationship between registered deaths and \texttt{WK} in adolescents and younger adults, indicates that fitting different smooths for males and for females is appropriate for \texttt{WK}.
The smooths are also clearly different across different age bands, so fitting different smooths for each age band is also indicated.

\begin{figure}
\centering
\includegraphics[width=0.99\textwidth]{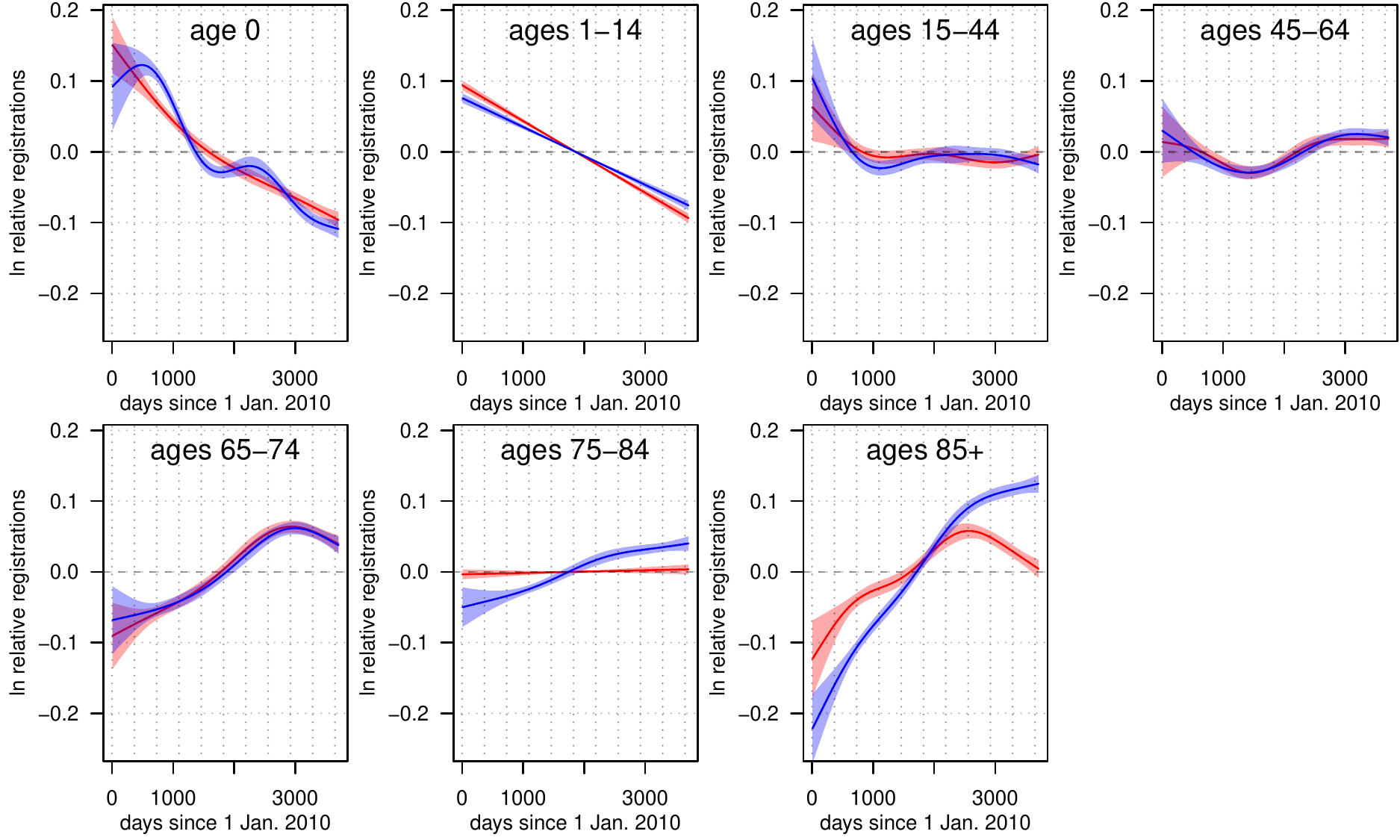}
\caption{Secular trends in registered deaths by age band and sex (red denoting females, blue males), measured on the GAM's natural-log.\ scale. Confidence bands extend $\pm 2$ standard errors around fitted trends. The $x$ axis is the date represented as the number of days since 1 Jan.\ 2010. Vertical dotted gridlines denote New Year's Day, 2010--2020.}
\label{fig:age-sex-wkeday}
\end{figure}

Figure \ref{fig:age-sex-wkeday} presents secular trends, fitted as smooths relating date to registered deaths for every age-sex combination.
There are undeniable differences in the trends with age; for example, registered deaths of young children steadily declined during the 2010s, but registered deaths of adults aged 85+ steadily increased from 2010 until 2017.
Within each under-75 age band, the smooth for males is similar to the smooth for females, but at ages 75--84 and especially for adults aged 85+ the male and female smooths are statistically and practically discrepant; the oldest men experienced a more relentless increase in registered deaths during the 2010s than the oldest women.
I therefore use a different secular-trend smooth for every combination of sex and age band in my main baseline-generating GAM.

\begin{figure}
\centering
\includegraphics[width=0.32\textwidth]{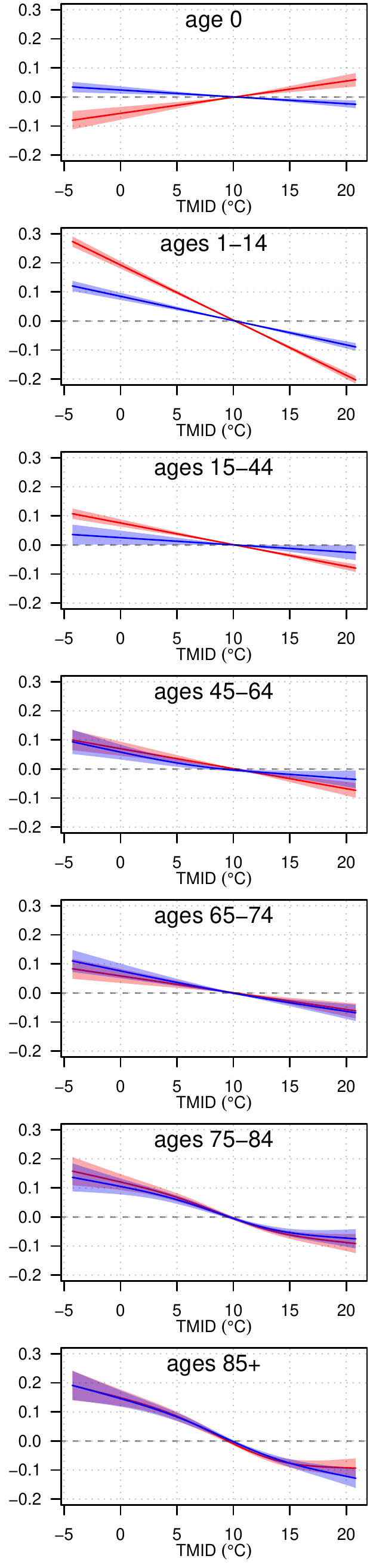}
\includegraphics[width=0.32\textwidth]{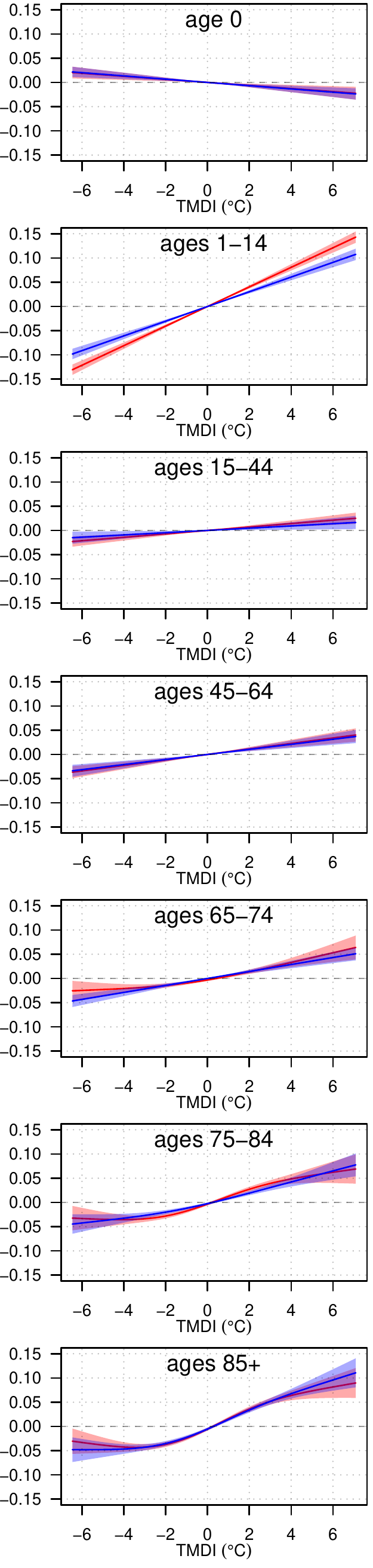}
\includegraphics[width=0.32\textwidth]{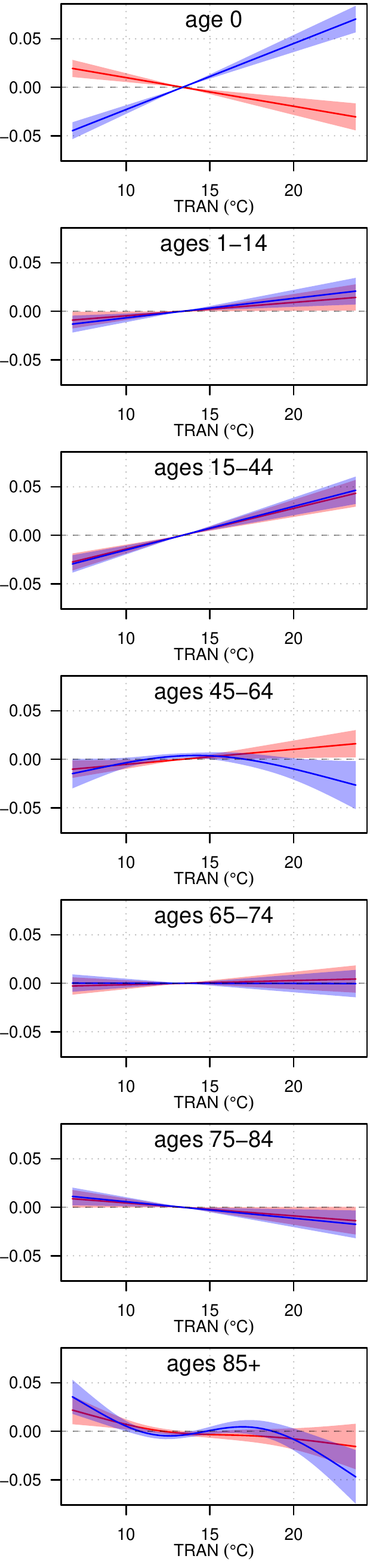}
\caption{Registered deaths as a function of \texttt{TMID} (left), \texttt{TMDI} (centre), and \texttt{TRAN} (right) by age band and sex (red denoting females and blue males), measured on the GAM's natural-log.\ scale. Confidence bands extend $\pm 2$ standard errors around each fit.}
\label{fig:age-sex-temperature}
\end{figure}

Figure \ref{fig:age-sex-temperature} presents the fitted smooths relating temperature to registered deaths for every age-sex combination.
Each smooth is limited to 3 DOF because it is implausible that mortality would have an intricate relationship to a given temperature variable.
In spite of the limited DOF, some of the smooths for older age bands are visibly nonlinear, although the \texttt{TRAN} smooths' deviations from nonlinearity are minute in absolute terms (at most 0.03 on the natural-log.\ scale) and unambiguously statistically significant only for men aged 85+.
In my main GAM, therefore, I fit only linear relationships between \texttt{TRAN} and registered deaths.
However, I fit a separate linear \texttt{TRAN}-mortality relationship for each age-sex combination, because the smooths in the full trial model differ substantially across age bands and, at age 0, differ significantly for boys versus girls.

I make a different simplification for \texttt{TMDI}.
\texttt{TMDI}'s smooths are substantially nonlinear for the oldest adults, so I continue using nonlinear smooths in the main GAM.
I also fit a different \texttt{TMDI} smooth for each age band, accounting for heterogeneity in smooths across age bands.
The male-female differences in smooths within each age band, by contrast, are statistically and practically insignificant for every age band except 1--14, and the difference in the 1--14 age band is practically insignificant (less than 0.04 on the natural-log.\ scale across the entire range of \texttt{TMDI} values).

I make no simplifications for \texttt{TMID}; that is, I fit 14 nonlinear smooths relating \texttt{TMID} to death registrations, one nonlinear smooth for each age-sex combination represented in the data.
In the full age-sex model, the \texttt{TMID} smooths have clear nonlinearities for older age bands, and multiple statistically and practically significant sex and age differences, indicating that the full complexity is necessary.

\begin{figure}
\centering
\includegraphics[width=0.99\textwidth]{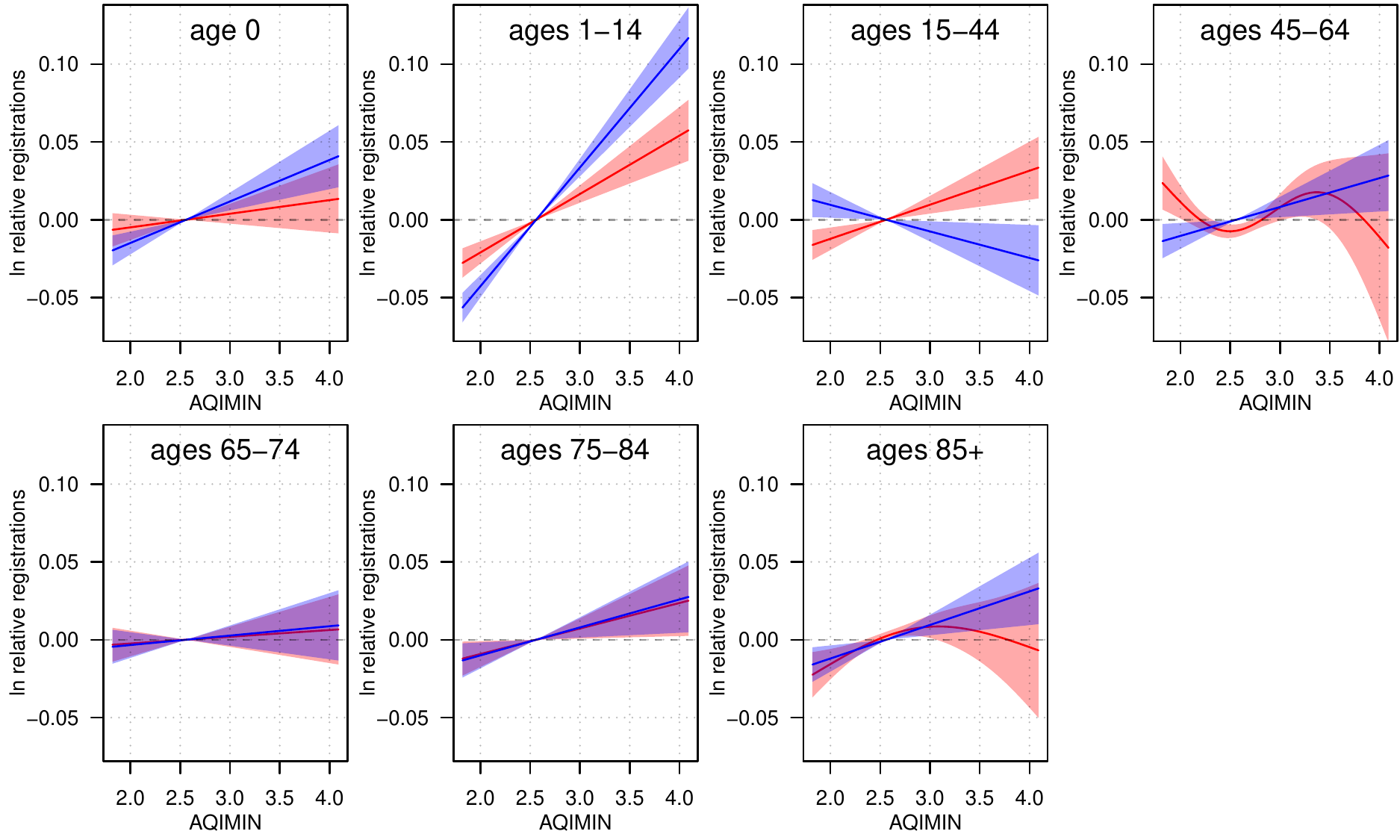}
\caption{Registered deaths as a function of \texttt{AQIMIN} by age band and sex (red denoting females and blue males), measured on the GAM's natural-log.\ scale. Confidence bands extend $\pm 2$ standard errors around each fit.}
\label{fig:age-sex-aqimin}
\end{figure}

The final set of smooths to review is that of \texttt{AQIMIN} (figure \ref{fig:age-sex-aqimin}).
Only two of the 14 fitted smooths --- those for women aged 45--64 and women aged 85+ --- are nonlinear.
One of those two (the smooth for women aged 85+) deviates from linearity to a statistically insignificant extent; the other deviates only minimally from linearity in absolute terms (at most 0.03 on the natural-log.\ scale).
Replacing the nonlinear smooths with linear fits in my main GAM is therefore an acceptable simplification.
On balance, I also decide that collapsing together sexes within each age band is also acceptable.
The male-female differences within age bands are consistently small, and while some of the differences are statistically significant, some of the sex-specific smooths have implausible negative gradients (implying decreasing mortality with worse air quality) which would be averaged out by using only one smooth per age band.
All in all, making both simplifications (substituting linear fits for nonlinear smooths, and using the same fit for males and females within an age band) is likely to give more sensible results than those produced by the full trial model which fits a different smooth for every age-sex combination (figure \ref{fig:age-sex-aqimin}).

This appendix's model simplifications reduce the time taken to fit the main GAM from 12.9 hours to 3.6 hours, and are of minimal detriment to the GAM's closeness of fit (the adjusted $R^2$ remained 0.992).

\newpage

\section{Estimates of excess registered deaths by sex}

\label{apx:age-week-sex-results}

The main text tabulates the weekly estimates of excess registered deaths by age band in table \ref{tab:excess-registrations}.
This appendix disaggregates the estimates by sex, in table \ref{tab:excess-registrations-by-sex}.
In both tables the ``$\pm$'' symbol denotes approximate standard errors of the GAM's baselines.
Here those standard errors are the GAM's sex-specific, root-mean-squared (RMS) log.-scale errors for the relevant week of the year in every year of the 2010s.
(The phrase ``log.-scale errors'' describes, for each week of the 2010s, the natural logarithm of the ratio of the actual number of deaths registered to the GAM's baseline.)

\begin{center}
\footnotesize
\begin{longtable}{crcrrrr}
\endfirsthead
\endhead
\toprule
& & & \multicolumn{3}{c}{all registered deaths} \\
\cmidrule{4-6}
ages & week & sex & GAM baseline & obs'd & excess \\
\midrule
\input{excess_registrations_by_sex.latex}
\bottomrule
\caption{\label{tab:excess-registrations-by-sex}Baseline versus observed counts of all registered deaths, by age band, week, and sex, registered in England and Wales in weeks 12--21 of 2020. ``$\pm$'' symbols denote approximate standard errors.}
\end{longtable}
\end{center}

\end{document}

%% file: excess_registrations.latex
%     0 & 12 & $  49 \pm   7$ &   44 & $  -5 \pm   7$ &  0 \\
% 1--14 & 12 & $  19 \pm   4$ &   12 & $  -7 \pm   4$ &  0 \\
15--44 & 12 & $ 285 \pm  18$ &  275 & $ -10 \pm  18$ &  1 \\
45--64 & 12 & $1251 \pm  41$ & 1264 & $  13 \pm  41$ &  6 \\
65--74 & 12 & $1717 \pm  63$ & 1780 & $  63 \pm  63$ & 20 \\
75--84 & 12 & $3028 \pm 111$ & 3067 & $  40 \pm 111$ & 31 \\
   85+ & 12 & $4237 \pm 221$ & 4204 & $ -33 \pm 221$ & 45 \\
\midrule
%     0 & 13 & $  50 \pm   7$ &   49 & $  -1 \pm   7$ &   0 \\
% 1--14 & 13 & $  19 \pm   4$ &   13 & $  -6 \pm   4$ &   0 \\
15--44 & 13 & $ 292 \pm  19$ &  283 & $ -10 \pm  19$ &   8 \\
45--64 & 13 & $1255 \pm  42$ & 1301 & $  46 \pm  42$ &  63 \\
65--74 & 13 & $1718 \pm  66$ & 1805 & $  87 \pm  66$ &  99 \\
75--84 & 13 & $3008 \pm  98$ & 3247 & $ 239 \pm  98$ & 181 \\
   85+ & 13 & $4191 \pm 192$ & 4444 & $ 253 \pm 192$ & 188 \\
\midrule
%     0 & 14 & $  49 \pm   7$ &   51 & $   2 \pm   7$ &    0 \\
% 1--14 & 14 & $  19 \pm   4$ &   21 & $   2 \pm   4$ &    0 \\
15--44 & 14 & $ 289 \pm  19$ &  288 & $  -1 \pm  19$ &   43 \\
45--64 & 14 & $1258 \pm  43$ & 1860 & $ 602 \pm  43$ &  412 \\
65--74 & 14 & $1732 \pm  68$ & 2734 & $1002 \pm  68$ &  626 \\
75--84 & 14 & $3080 \pm 105$ & 5005 & $1925 \pm 105$ & 1231 \\
   85+ & 14 & $4306 \pm 201$ & 6428 & $2122 \pm 201$ & 1163 \\
\midrule
%     0 & 15 & $  43 \pm   8$ &   38 & $  -5 \pm   8$ &    0 \\
% 1--14 & 15 & $  17 \pm   4$ &   14 & $  -3 \pm   4$ &    0 \\
15--44 & 15 & $ 260 \pm  22$ &  332 & $  72 \pm  22$ &   74 \\
45--64 & 15 & $1100 \pm  50$ & 2111 & $1011 \pm  50$ &  742 \\
65--74 & 15 & $1534 \pm  76$ & 2946 & $1412 \pm  76$ & 1104 \\
75--84 & 15 & $2634 \pm 135$ & 5613 & $2980 \pm 135$ & 2210 \\
   85+ & 15 & $3684 \pm 264$ & 7462 & $3778 \pm 264$ & 2083 \\
\midrule
%     0 & 16 & $  48 \pm   9$ &   51 & $   2 \pm   9$ &    0 \\
% 1--14 & 16 & $  17 \pm   5$ &   15 & $  -2 \pm   5$ &    2 \\
15--44 & 16 & $ 287 \pm  25$ &  353 & $  66 \pm  25$ &  101 \\
45--64 & 16 & $1179 \pm  57$ & 2294 & $1115 \pm  57$ &  966 \\
65--74 & 16 & $1615 \pm  76$ & 3380 & $1765 \pm  76$ & 1442 \\
75--84 & 16 & $2740 \pm 144$ & 6657 & $3917 \pm 144$ & 2834 \\
   85+ & 16 & $3729 \pm 304$ & 9601 & $5872 \pm 304$ & 3413 \\
\midrule
%     0 & 17 & $  55 \pm   8$ &   54 & $  -1 \pm   8$ &    0 \\
% 1--14 & 17 & $  20 \pm   5$ &   12 & $  -8 \pm   5$ &    0 \\
15--44 & 17 & $ 326 \pm  25$ &  404 & $  78 \pm  25$ &  103 \\
45--64 & 17 & $1369 \pm  57$ & 2283 & $ 914 \pm  57$ &  823 \\
65--74 & 17 & $1891 \pm  65$ & 3238 & $1347 \pm  65$ & 1189 \\
75--84 & 17 & $3262 \pm 113$ & 6513 & $3251 \pm 113$ & 2615 \\
   85+ & 17 & $4485 \pm 260$ & 9493 & $5008 \pm 260$ & 3507 \\
\midrule
%     0 & 18 & $  50 \pm   7$ &   48 & $  -2 \pm   7$ &    0 \\
% 1--14 & 18 & $  18 \pm   4$ &   11 & $  -7 \pm   4$ &    0 \\
15--44 & 18 & $ 292 \pm  22$ &  345 & $  52 \pm  22$ &   52 \\
45--64 & 18 & $1233 \pm  49$ & 1897 & $ 664 \pm  49$ &  512 \\
65--74 & 18 & $1703 \pm  51$ & 2601 & $ 898 \pm  51$ &  805 \\
75--84 & 18 & $2938 \pm  85$ & 5142 & $2204 \pm  85$ & 1866 \\
   85+ & 18 & $3982 \pm 176$ & 7909 & $3927 \pm 176$ & 2800 \\
\midrule
%     0 & 19 & $  43 \pm   7$ &   28 & $ -15 \pm   7$ &    1 \\
% 1--14 & 19 & $  16 \pm   4$ &   20 & $   4 \pm   4$ &    0 \\
15--44 & 19 & $ 258 \pm  24$ &  233 & $ -25 \pm  24$ &   40 \\
45--64 & 19 & $1076 \pm  51$ & 1370 & $ 294 \pm  51$ &  317 \\
65--74 & 19 & $1488 \pm  59$ & 1935 & $ 447 \pm  59$ &  486 \\
75--84 & 19 & $2550 \pm 103$ & 3627 & $1076 \pm 103$ & 1236 \\
   85+ & 19 & $3476 \pm 149$ & 5444 & $1968 \pm 149$ & 1850 \\
\midrule
%     0 & 20 & $  52 \pm   6$ &   56 & $   4 \pm   6$ &    1 \\
% 1--14 & 20 & $  19 \pm   3$ &   19 & $   0 \pm   3$ &    0 \\
15--44 & 20 & $ 309 \pm  19$ &  287 & $ -22 \pm  19$ &   33 \\
45--64 & 20 & $1261 \pm  41$ & 1643 & $ 382 \pm  41$ &  266 \\
65--74 & 20 & $1733 \pm  56$ & 2188 & $ 455 \pm  56$ &  485 \\
75--84 & 20 & $2938 \pm 106$ & 4167 & $1229 \pm 106$ & 1155 \\
   85+ & 20 & $3936 \pm 107$ & 6213 & $2277 \pm 107$ & 1870 \\
\midrule
%     0 & 21 & $  49 \pm  11$ &   51 & $   2 \pm  11$ &    0 \\
% 1--14 & 21 & $  20 \pm   6$ &   16 & $  -4 \pm   6$ &    0 \\
15--44 & 21 & $ 295 \pm  33$ &  339 & $  44 \pm  33$ &   30 \\
45--64 & 21 & $1240 \pm  68$ & 1481 & $ 241 \pm  68$ &  198 \\
65--74 & 21 & $1728 \pm 100$ & 1883 & $ 155 \pm 100$ &  302 \\
75--84 & 21 & $2920 \pm 186$ & 3455 & $ 536 \pm 186$ &  773 \\
   85+ & 21 & $3948 \pm 190$ & 5063 & $1115 \pm 190$ & 1286 \\

%% file: excess_registrations_by_sex.latex
     0 & 12 & female & $  21 \pm   6$ &   19 & $  -2 \pm   6$ \\
     0 & 12 &  male  & $  28 \pm   8$ &   25 & $  -3 \pm   8$ \\
 1--14 & 12 & female & $   8 \pm   4$ &    4 & $  -4 \pm   4$ \\
 1--14 & 12 &  male  & $  10 \pm   5$ &    8 & $  -2 \pm   5$ \\
15--44 & 12 & female & $ 104 \pm  10$ &   90 & $ -14 \pm  10$ \\
15--44 & 12 &  male  & $ 181 \pm  17$ &  185 & $   4 \pm  17$ \\
45--64 & 12 & female & $ 504 \pm  22$ &  497 & $  -7 \pm  22$ \\
45--64 & 12 &  male  & $ 747 \pm  33$ &  767 & $  20 \pm  33$ \\
65--74 & 12 & female & $ 719 \pm  28$ &  747 & $  28 \pm  28$ \\
65--74 & 12 &  male  & $ 998 \pm  39$ & 1033 & $  35 \pm  39$ \\
75--84 & 12 & female & $1418 \pm  53$ & 1412 & $  -6 \pm  53$ \\
75--84 & 12 &  male  & $1610 \pm  60$ & 1655 & $  45 \pm  60$ \\
   85+ & 12 & female & $2541 \pm 136$ & 2479 & $ -62 \pm 136$ \\
   85+ & 12 &  male  & $1695 \pm  91$ & 1725 & $  30 \pm  91$ \\
\midrule
     0 & 13 & female & $  21 \pm   5$ &   22 & $   1 \pm   5$ \\
     0 & 13 &  male  & $  29 \pm   6$ &   27 & $  -2 \pm   6$ \\
 1--14 & 13 & female & $   8 \pm   3$ &    8 & $   0 \pm   3$ \\
 1--14 & 13 &  male  & $  11 \pm   4$ &    5 & $  -6 \pm   4$ \\
15--44 & 13 & female & $ 107 \pm  12$ &   97 & $ -10 \pm  12$ \\
15--44 & 13 &  male  & $ 185 \pm  20$ &  186 & $   1 \pm  20$ \\
45--64 & 13 & female & $ 506 \pm  26$ &  537 & $  31 \pm  26$ \\
45--64 & 13 &  male  & $ 748 \pm  39$ &  764 & $  16 \pm  39$ \\
65--74 & 13 & female & $ 719 \pm  37$ &  716 & $  -3 \pm  37$ \\
65--74 & 13 &  male  & $ 999 \pm  51$ & 1089 & $  90 \pm  51$ \\
75--84 & 13 & female & $1405 \pm  58$ & 1454 & $  49 \pm  58$ \\
75--84 & 13 &  male  & $1603 \pm  66$ & 1793 & $ 190 \pm  66$ \\
   85+ & 13 & female & $2516 \pm 139$ & 2577 & $  61 \pm 139$ \\
   85+ & 13 &  male  & $1676 \pm  93$ & 1867 & $ 191 \pm  93$ \\
\midrule
     0 & 14 & female & $  21 \pm   5$ &   25 & $   4 \pm   5$ \\
     0 & 14 &  male  & $  28 \pm   6$ &   26 & $  -2 \pm   6$ \\
 1--14 & 14 & female & $   8 \pm   3$ &   14 & $   6 \pm   3$ \\
 1--14 & 14 &  male  & $  11 \pm   4$ &    7 & $  -4 \pm   4$ \\
15--44 & 14 & female & $ 105 \pm  12$ &  105 & $   0 \pm  12$ \\
15--44 & 14 &  male  & $ 184 \pm  22$ &  183 & $  -1 \pm  22$ \\
45--64 & 14 & female & $ 505 \pm  23$ &  703 & $ 198 \pm  23$ \\
45--64 & 14 &  male  & $ 753 \pm  35$ & 1157 & $ 404 \pm  35$ \\
65--74 & 14 & female & $ 721 \pm  35$ & 1033 & $ 312 \pm  35$ \\
65--74 & 14 &  male  & $1010 \pm  48$ & 1701 & $ 691 \pm  48$ \\
75--84 & 14 & female & $1431 \pm  57$ & 2132 & $ 701 \pm  57$ \\
75--84 & 14 &  male  & $1649 \pm  65$ & 2873 & $1224 \pm  65$ \\
   85+ & 14 & female & $2586 \pm 129$ & 3581 & $ 995 \pm 129$ \\
   85+ & 14 &  male  & $1719 \pm  86$ & 2847 & $1128 \pm  86$ \\
\midrule
     0 & 15 & female & $  18 \pm   5$ &   10 & $  -8 \pm   5$ \\
     0 & 15 &  male  & $  25 \pm   7$ &   28 & $   3 \pm   7$ \\
 1--14 & 15 & female & $   8 \pm   2$ &    7 & $   0 \pm   2$ \\
 1--14 & 15 &  male  & $  10 \pm   3$ &    7 & $  -3 \pm   3$ \\
15--44 & 15 & female & $  93 \pm   8$ &  131 & $  38 \pm   8$ \\
15--44 & 15 &  male  & $ 167 \pm  14$ &  201 & $  34 \pm  14$ \\
45--64 & 15 & female & $ 442 \pm  27$ &  796 & $ 354 \pm  27$ \\
45--64 & 15 &  male  & $ 658 \pm  41$ & 1315 & $ 656 \pm  41$ \\
65--74 & 15 & female & $ 641 \pm  34$ & 1142 & $ 501 \pm  34$ \\
65--74 & 15 &  male  & $ 894 \pm  47$ & 1804 & $ 910 \pm  47$ \\
75--84 & 15 & female & $1221 \pm  64$ & 2367 & $1146 \pm  64$ \\
75--84 & 15 &  male  & $1412 \pm  74$ & 3246 & $1834 \pm  74$ \\
   85+ & 15 & female & $2195 \pm 153$ & 4115 & $1920 \pm 153$ \\
   85+ & 15 &  male  & $1489 \pm 104$ & 3347 & $1858 \pm 104$ \\
\midrule
     0 & 16 & female & $  20 \pm   4$ &   27 & $   7 \pm   4$ \\
     0 & 16 &  male  & $  28 \pm   6$ &   24 & $  -4 \pm   6$ \\
 1--14 & 16 & female & $   8 \pm   3$ &    8 & $   0 \pm   3$ \\
 1--14 & 16 &  male  & $  10 \pm   3$ &    7 & $  -3 \pm   3$ \\
15--44 & 16 & female & $ 102 \pm  11$ &  136 & $  34 \pm  11$ \\
15--44 & 16 &  male  & $ 184 \pm  19$ &  217 & $  33 \pm  19$ \\
45--64 & 16 & female & $ 476 \pm  28$ &  893 & $ 417 \pm  28$ \\
45--64 & 16 &  male  & $ 702 \pm  42$ & 1401 & $ 698 \pm  42$ \\
65--74 & 16 & female & $ 672 \pm  39$ & 1309 & $ 637 \pm  39$ \\
65--74 & 16 &  male  & $ 943 \pm  55$ & 2071 & $1128 \pm  55$ \\
75--84 & 16 & female & $1267 \pm  70$ & 2922 & $1655 \pm  70$ \\
75--84 & 16 &  male  & $1472 \pm  81$ & 3735 & $2263 \pm  81$ \\
   85+ & 16 & female & $2218 \pm 167$ & 5611 & $3393 \pm 167$ \\
   85+ & 16 &  male  & $1511 \pm 114$ & 3990 & $2479 \pm 114$ \\
\midrule
     0 & 17 & female & $  23 \pm   5$ &   21 & $  -2 \pm   5$ \\
     0 & 17 &  male  & $  31 \pm   7$ &   33 & $   2 \pm   7$ \\
 1--14 & 17 & female & $   9 \pm   2$ &    3 & $  -6 \pm   2$ \\
 1--14 & 17 &  male  & $  12 \pm   3$ &    9 & $  -3 \pm   3$ \\
15--44 & 17 & female & $ 115 \pm  11$ &  153 & $  38 \pm  11$ \\
15--44 & 17 &  male  & $ 210 \pm  20$ &  251 & $  40 \pm  20$ \\
45--64 & 17 & female & $ 551 \pm  25$ &  849 & $ 298 \pm  25$ \\
45--64 & 17 &  male  & $ 818 \pm  36$ & 1434 & $ 616 \pm  36$ \\
65--74 & 17 & female & $ 784 \pm  35$ & 1223 & $ 439 \pm  35$ \\
65--74 & 17 &  male  & $1107 \pm  49$ & 2015 & $ 908 \pm  49$ \\
75--84 & 17 & female & $1504 \pm  65$ & 2892 & $1388 \pm  65$ \\
75--84 & 17 &  male  & $1758 \pm  76$ & 3621 & $1863 \pm  76$ \\
   85+ & 17 & female & $2665 \pm 192$ & 5628 & $2963 \pm 192$ \\
   85+ & 17 &  male  & $1820 \pm 131$ & 3865 & $2045 \pm 131$ \\
\midrule
     0 & 18 & female & $  21 \pm   4$ &   18 & $  -3 \pm   4$ \\
     0 & 18 &  male  & $  28 \pm   5$ &   30 & $   2 \pm   5$ \\
 1--14 & 18 & female & $   8 \pm   2$ &    6 & $  -2 \pm   2$ \\
 1--14 & 18 &  male  & $  10 \pm   3$ &    5 & $  -5 \pm   3$ \\
15--44 & 18 & female & $ 104 \pm  15$ &  132 & $  28 \pm  15$ \\
15--44 & 18 &  male  & $ 189 \pm  26$ &  213 & $  24 \pm  26$ \\
45--64 & 18 & female & $ 498 \pm  26$ &  714 & $ 216 \pm  26$ \\
45--64 & 18 &  male  & $ 735 \pm  38$ & 1183 & $ 448 \pm  38$ \\
65--74 & 18 & female & $ 705 \pm  23$ & 1030 & $ 325 \pm  23$ \\
65--74 & 18 &  male  & $ 998 \pm  32$ & 1571 & $ 573 \pm  32$ \\
75--84 & 18 & female & $1354 \pm  50$ & 2382 & $1028 \pm  50$ \\
75--84 & 18 &  male  & $1584 \pm  58$ & 2760 & $1176 \pm  58$ \\
   85+ & 18 & female & $2368 \pm 114$ & 4814 & $2446 \pm 114$ \\
   85+ & 18 &  male  & $1614 \pm  77$ & 3095 & $1481 \pm  77$ \\
\midrule
     0 & 19 & female & $  18 \pm   5$ &   12 & $  -6 \pm   5$ \\
     0 & 19 &  male  & $  25 \pm   7$ &   16 & $  -9 \pm   7$ \\
 1--14 & 19 & female & $   7 \pm   3$ &    7 & $   0 \pm   3$ \\
 1--14 & 19 &  male  & $   9 \pm   4$ &   13 & $   4 \pm   4$ \\
15--44 & 19 & female & $  92 \pm  12$ &   86 & $  -6 \pm  12$ \\
15--44 & 19 &  male  & $ 166 \pm  22$ &  147 & $ -19 \pm  22$ \\
45--64 & 19 & female & $ 436 \pm  22$ &  523 & $  86 \pm  22$ \\
45--64 & 19 &  male  & $ 640 \pm  33$ &  847 & $ 208 \pm  33$ \\
65--74 & 19 & female & $ 618 \pm  31$ &  806 & $ 188 \pm  31$ \\
65--74 & 19 &  male  & $ 870 \pm  44$ & 1129 & $ 259 \pm  44$ \\
75--84 & 19 & female & $1178 \pm  50$ & 1648 & $ 470 \pm  50$ \\
75--84 & 19 &  male  & $1372 \pm  59$ & 1979 & $ 606 \pm  59$ \\
   85+ & 19 & female & $2064 \pm  95$ & 3366 & $1302 \pm  95$ \\
   85+ & 19 &  male  & $1413 \pm  65$ & 2078 & $ 665 \pm  65$ \\
\midrule
     0 & 20 & female & $  21 \pm   4$ &   22 & $   1 \pm   4$ \\
     0 & 20 &  male  & $  31 \pm   6$ &   34 & $   3 \pm   6$ \\
 1--14 & 20 & female & $   8 \pm   2$ &    8 & $   0 \pm   2$ \\
 1--14 & 20 &  male  & $  11 \pm   2$ &   11 & $   0 \pm   2$ \\
15--44 & 20 & female & $ 112 \pm  13$ &  107 & $  -5 \pm  13$ \\
15--44 & 20 &  male  & $ 197 \pm  23$ &  180 & $ -17 \pm  23$ \\
45--64 & 20 & female & $ 517 \pm  26$ &  656 & $ 139 \pm  26$ \\
45--64 & 20 &  male  & $ 745 \pm  37$ &  987 & $ 242 \pm  37$ \\
65--74 & 20 & female & $ 721 \pm  30$ &  898 & $ 177 \pm  30$ \\
65--74 & 20 &  male  & $1012 \pm  43$ & 1290 & $ 278 \pm  43$ \\
75--84 & 20 & female & $1361 \pm  58$ & 1874 & $ 513 \pm  58$ \\
75--84 & 20 &  male  & $1578 \pm  68$ & 2293 & $ 716 \pm  68$ \\
   85+ & 20 & female & $2339 \pm  53$ & 3765 & $1426 \pm  53$ \\
   85+ & 20 &  male  & $1597 \pm  36$ & 2448 & $ 851 \pm  36$ \\
\midrule
     0 & 21 & female & $  21 \pm   5$ &   25 & $   4 \pm   5$ \\
     0 & 21 &  male  & $  28 \pm   6$ &   26 & $  -2 \pm   6$ \\
 1--14 & 21 & female & $   8 \pm   2$ &    4 & $  -4 \pm   2$ \\
 1--14 & 21 &  male  & $  11 \pm   3$ &   12 & $   1 \pm   3$ \\
15--44 & 21 & female & $ 106 \pm  16$ &  117 & $  12 \pm  16$ \\
15--44 & 21 &  male  & $ 189 \pm  28$ &  222 & $  33 \pm  28$ \\
45--64 & 21 & female & $ 503 \pm  27$ &  575 & $  72 \pm  27$ \\
45--64 & 21 &  male  & $ 738 \pm  39$ &  906 & $ 168 \pm  39$ \\
65--74 & 21 & female & $ 721 \pm  57$ &  759 & $  38 \pm  57$ \\
65--74 & 21 &  male  & $1006 \pm  80$ & 1124 & $ 118 \pm  80$ \\
75--84 & 21 & female & $1351 \pm 104$ & 1633 & $ 282 \pm 104$ \\
75--84 & 21 &  male  & $1568 \pm 121$ & 1822 & $ 254 \pm 121$ \\
   85+ & 21 & female & $2343 \pm 175$ & 3078 & $ 735 \pm 175$ \\
   85+ & 21 &  male  & $1606 \pm 120$ & 1985 & $ 379 \pm 120$ \\